\documentclass{article}

\usepackage[margin=1in]{geometry}
\usepackage{graphicx}
\usepackage[colorlinks=true, citecolor=blue, linkcolor=blue]{hyperref}
\usepackage{amsmath}
\usepackage{amssymb}
\usepackage{bm}

\usepackage[backend=bibtex,style=numeric,sorting=none, natbib=true]{biblatex}
\usepackage{xr}

\usepackage{savesym}
\savesymbol{tablenum}
\usepackage[separate-uncertainty=true,
multi-part-units=single, 
print-unity-mantissa=false]{siunitx}
\restoresymbol{SIX}{tablenum}

\DeclareSIUnit\year{yr}

\usepackage{xargs}

\usepackage{xspace}
\newcommand{\Kporo}{K23\@\xspace}
\newcommand{\RNporo}{RN22\@\xspace}

\newcommand{\Grad}{{\boldsymbol{\nabla}}}
\newcommand{\Div}{\Grad\cdot}
\newcommand{\Divb}[1]{\Grad\cdot\brackets{#1}}
\newcommand{\Gradb}[1]{\Grad\brackets{#1}}
\newcommand{\Gradr}[1]{\partial_r {#1}}

\newcommand{\Gradtheta}[1]{\partial_\theta}
\newcommand{\Gradphi}[1]{\partial_\phi}

\newcommand{\rhat}{\boldsymbol{e}_r}
\newcommand{\thetahat}{\hat{\boldsymbol{e}}_{\theta}}
\newcommand{\phihat}{\hat{\boldsymbol{e}}_{\varphi}}

\newcommand{\porosity}{\phi}

\newcommand{\brackets}[1]{\left(#1\right)}

\usepackage{xargs}

\newcommandx{\nm}[3][2=,3=]{#1_{n #2}^{m #3}}
\newcommandx{\nmd}[3][2=,3=]{#1_{n' #2}^{m' #3}}
\newcommandx{\sphY}[3]{#1_{#2}^{#3}}

\renewcommand{\vec}{\bm}

\newcommand{\stressT}{\vec{\sigma}}

\newcommand{\strainT}{\vec{\epsilon}}

\newcommand{\trace}[1]{\text{tr}\brackets{#1}}
\newcommand{\mat}[1]{\bm{#1}}

\newcommand{\identity}{\mat{1}}

\usepackage{etoolbox}%

\newcommand{\pos}{\vec{r}}

\newcommand{\disS}{\vec{u}_s}
\newcommand{\disL}{\vec{u}_l}
\newcommand{\disR}{\vec{u}_\text{rel}}
\newcommand{\dis}{\vec{u}}

\newcommand{\f}[1]{\tilde{#1}}
\newcommand{\s}[1]{\hat{#1}}
\newcommand{\den}{\rho}
\newcommand{\pot}{\Phi}

\newcommand{\clat}{\theta}
\newcommand{\lon}{\varphi}

\newcommand{\Rnm}{\vec{R}_n^m}
\newcommand{\Snm}{\vec{S}_n^m}
\newcommand{\Ynm}{Y_n^m}

\newcommand{\denS}[1]{%
	\ifstrequal{#1}{}%
	{\den_s}{\den_{s,#1}}
}

\newcommand{\denL}[1]{%
	\ifstrequal{#1}{}%
	{\den_l}{\den_{l,#1}}
}

\renewcommand{\Im}[1]{\text{Im}\left( #1 \right)}

\begin{document}

\title{Poro-viscoelastic tidal heating of Io}

\author{
	Hamish C. F. C. Hay$^{1,4*}$, Ian Hewitt$^{2}$, Marc Rovira-Navarro$^{3}$,
	Richard F. Katz$^{1}$}




\date{}

\maketitle
\vspace{-2cm}
\begin{center}
	\section*{}
	\small$^{1}$Department of Earth Sciences, University of Oxford, Oxford, UK.\\
	\small$^{2}$Mathematical Institute, University of Oxford, Oxford, UK.\\
	\small$^{3}$Faculty of Aerospace Engineering, TU Delft, Delft, The Netherlands.\\
	\small$^{4}$Now at: School of Mathematics and Statistics, University of St Andrews, St Andrews, UK.\\
	\small$^\ast$Corresponding author. Email: hcfch1@st-andrews.ac.uk
\end{center}

\begin{abstract}
	Io's tidally driven global volcanism indicates widespread partial melting in its mantle. 
	How this melt participates in the interior dynamics, and, in particular, the role it plays in tidal dissipation, is poorly understood. 
	We model Io's tidal deformation by treating its mantle as a two-phase (solid and melt) system. 
	By combining poro-viscous and poro-elastic compaction theories in a Maxwell framework with a consistent model of tidal and self-gravitation, we produce the first self-consistent evaluation of Io's tidal heating rate due to shearing, compaction, and Darcy flow.

	We find that Darcy dissipation can potentially exceed shear heating, but only for large (0.05 to 0.2) melt fractions, and if the grain size is large or melt viscosity ultra-low. 
	Since grain sizes larger than 1~cm are unlikely, this suggests that Darcy dissipation is secondary to shear dissipation. Compaction dissipation is maximised when the asthenosphere is highly resistive to isotropic stresses, but contributes at most 1\% of Io's observed heating rate. 
	
	This work represents a crucial step toward a self-consistent quantitative theory for the dynamics of Io's partially molten interior.
\end{abstract}

\maketitle

\section{Introduction}
\label{sec:intro}

Io, the innermost Galilean moon of Jupiter, is the most volcanically active body in our solar system \cite{mcewen2002ActiveVolcanism}, with a thermal output of $\sim$\SI{100}{T\watt} \cite{lainey2009StrongTidal}, roughly $10^5$ times that of Earth's similarly sized moon \cite{williams2015TidesMoon}. Io's volcanism is powered by tides, which transfer orbital energy to the interior via Jupiter's gravitational field \cite{peale1979MeltingIo}. These tides are the result of viscoelastic deformation in its interior, dissipating heat that drives melting. The buoyant melt segregates from the residual solid, accumulating at the near-surface before either refreezing or erupting \cite{moore2001ThermalState}. As melt segregates, the residual solid compacts \cite{mckenzie1984GenerationCompaction}. A physically consistent description of Io's internal dynamics must therefore combine tidal deformation, melt segregation, compaction, and viscous dissipation into a single theory describing the two-phase dynamics of partially molten rock. Here we take a first step towards this theory.

Io's volcanism has been observed for decades with passing spacecraft \cite{morabito1979DiscoveryCurrently, keszthelyi2001ImagingVolcanic, spencer2007IoVolcanism, mura2020InfraredObservations} and ground-based telescopes \cite{witteborn1979IoIntense,veeder1994IosHeat, spencer1997PelePlume, rathbun2010GroundbasedObservations, dekleer2019VariabilityIo, pater20212020Observational}. The texture and internal location of the erupted magma's reservoir has remained elusive.
The detection of an anomalous, induced magnetic field around Io \cite{khurana2011EvidenceGlobal} suggested the presence of an electrically conductive subsurface layer, consistent with an internal magma ocean. However, this interpretation has been disfavoured by subsequent reanalysis of the induced magnetic-field data \cite{blocker2018MHDModeling}, observations of Io's aurora \cite{roth2017ConstraintsIos} and, most conclusively, by the gravitational signature of Io's tidal deformation measured by the Juno spacecraft \cite{park2024IosTidal}. A significant fraction of Io's mantle is therefore partially molten, a state in which the rock comprises two phases: a porous skeleton of solid grains interpenetrated by liquid melt. 

In partially molten material (e.g., silicate rock), melt segregates by porous flow, driven by pressure gradients and body forces. This is described by a modified Darcy's law. The long-term, creeping deformation of the residual solid is viscous, including its compaction in response to melt segregation. This is described by a modified Stokes force-balance equation. The coupling of the Darcy and Stokes equations is referred to as poro-viscous compaction theory \cite{mckenzie1984GenerationCompaction, katz2022DynamicsPartially}. An associated rheological model supplies viscosities for deviatoric (shear) and isotropic (compaction) viscous deformation.  This theory has long been applied to partially molten regions of Earth's mantle \cite{spiegelman1993physics,katz2022physics}, and also to terrestrial ice \cite{fowler1984transport, schoof2016model, haseloff2019englacial}.  Distinct from the terrestrial case, however, the energy that melts Io's interior is that dissipated as heat by tides \cite{peale1979MeltingIo}. Recent work considers the coupled melting, melt segregation and viscous compaction of Io's mantle \cite{spencer2020CompositionalLayering, spencer2020MagmaticIntrusions, spencer2021TidalControls, miyazaki2022SubsurfaceMagma}, but no work has self-consistently coupled tidal deformation and heat generation with melting \cite{steinke2020TidallyInduced}. 

Friction associated with Io's repeated tidal distortion releases heat as Jupiter's gravity field periodically deforms the body into a triaxial figure. Tidal heating has classically been modelled by treating Io's mantle with viscoelastic dissipation theory \cite{segatz1988TidalDissipation}, which partially stems back to the classic work of A.E.~Love \cite{love1911ProblemsGeodynamics}. A key prediction of such modelling is that tidal heat input is spatially variable, with different heat distributions possible depending on internal structure \cite{beuthe2013SpatialPatterns}. Melting must also therefore be spatially variable. Melt lubricates grain boundaries, lowering the shear viscosity of the mantle \cite{bierson2016TestIo}. This means that regions of high melt fraction have lower viscosity and are more susceptible to tidal heating \cite{veenstra2025LateralMelta}. There is thus an inherent coupling between melt and tidal heat generation. Furthermore, tidally induced deformation of Io's mantle drives melt segregation through (de)compaction. This segregation alters Io's tidal response and drives heating through Darcy flow and viscous compaction. Ref.~\cite{kervazo2021SolidTides} models the latter process, but treats Io's mantle as a single-phase (solid) material. Hence, ref.~\cite{kervazo2021SolidTides}'s approach cannot self-consistently take into account melt segregation. A better approach to closing (and understanding) Io's heat budget, is therefore to combine poro-viscous and viscoelastic theories. 

The first poro-viscoelastic model of tidal deformation was published in ref.~\cite{liao2020HeatProduction} to quantify heating in the porous core of Saturn's icy moon, Enceladus. That work reduced the problem to a kinematic one, whereby the two-phase core was forced by imposed displacements its surface. Later, ref.~\cite{rovira-navarro2022TidesEnceladus} improved upon this by imposing appropriate boundary conditions, forcing the core with the tidal potential, and coupling fluid displacements to the gravity field. Most recently, ref.~\cite{kamata2023PoroviscoelasticGravitational} provided a more comprehensive derivation of poro-viscoelastic gravitational dynamics, including detailed consideration of the problem in several limiting cases, noting inconsistencies in both refs.~\cite{liao2020HeatProduction} and \cite{rovira-navarro2022TidesEnceladus}, and considering radial porosity variations. There are, however, two main deficiencies of all of these works. Firstly, viscous compaction is neglected, despite the fact that it is relevant at the same time-scales as viscous shear.  Secondly, rheological properties are assumed to be laterally uniform (also a common assumption in one-phase tidal deformation models). For the reasons outlined above, a theory of Io that can plausibly address the ensemble of observations must include both of these effects. Toward this end, here we incorporate viscous compaction while leaving laterally varying properties for future work. 

We model the tidal deformation of Io by treating its interior as a poro-viscoelastic continuum and solving the deformation equations of a self-gravitating body \cite{segatz1988TidalDissipation}, described by the two-phase theory developed in refs.~\cite{rovira-navarro2022TidesEnceladus, kamata2023PoroviscoelasticGravitational}. We extend the theory by (\textit{i}) modifying the isotropic component of the rheological model to self-consistently account for isotropic  viscous deformation, and (\textit{ii}) formulating the constitutive laws in a physically and mathematically consistent manner based on the original work of Love \cite{love1911ProblemsGeodynamics}. We treat Io's asthenosphere as a two-phase, Maxwell material with impermeable boundaries above and below, exploring how tidal forcing circulates melt in the asthenosphere and dissipates heat due to solid deformation (deviatoric and isotropic) as well as porous flow. We find that dissipation due to (de)compaction can be at most 1--10\% of Io's observed heating rate, but the upper end of this range requires an effective grain size $\gtrsim$~\SI{10}{\cm}. We also find that dissipation from porous flow can generate substantial heating only for a highly permeable asthenosphere, and this state may be possible if, again, the mantle's effective grain size is $\gtrsim$~\SI{10}{\cm}.

The manuscript is organised as follows. In \S\ref{sec:theory}, we describe the theory of poro-viscoelasticity in a compacting, self-gravitating body, and our approach to the problem. The relevant internal structures and material parameters of Io are outlined in \S\ref{sec:internal_structure}, and corresponding tidal deformation and heating results are given in \S\ref{sec:results}. We summarise the potential next steps for this work in \S\ref{sec:discussion}, before concluding in \S\ref{sec:conclusion}.

\section{Theory} \label{sec:theory}

Here we recap the theory of two-phase mechanics of a self-gravitating, poro-viscoelastic planetary body. The theory builds on the recent developments by refs.~\cite{liao2020HeatProduction, rovira-navarro2022TidesEnceladus, kamata2023PoroviscoelasticGravitational}; the latter two papers are hereafter \RNporo and \Kporo, respectively. We largely formulate the problem following \Kporo, but base the rheological formalism on the original arguments of Love \cite{love1911ProblemsGeodynamics} and incorporate a compaction viscosity \cite{mckenzie1984GenerationCompaction}. We show that Love's original formulation of the problem provides an intuitive, natural extension to poro-viscoelasticity that avoids a mathematical ambiguity often present in the literature. 

In \S\ref{ssec:gov}, we present the governing equations of poro-viscoelastic gravitational dynamics, followed by their linearisation and Fourier transform in \S\ref{ssec:theory_linearise}. Calculation of dissipation from shear, compaction, and Darcy deformation is described in \S\ref{ssec:theory_dissipation}. 
\subsection{Governing Equations}\label{ssec:gov}

In the following, we denote quantities relevant to individual phases with the subscripts $l$ (liquid) and $s$ (solid).
When multiple phases are present within a layer, each individual phase requires its own set of conservation laws. For mass conservation these are, 
\begin{subequations}\label{eq:gov_mass}
	\begin{align}
		\frac{\partial \brackets{\porosity \rho_l} }{\partial t} +\Divb{\porosity \rho_l \bm{v}_l} &= \Gamma, \label{eq:gov_mass_l} \\
		\frac{\partial \brackets{\brackets{1 - \porosity} \rho_s}}{\partial t} +\Divb{\brackets{1 - \porosity} \rho_s \bm{v}_s} &= -\Gamma, \label{eq:gov_mass_s}
	\end{align}
\end{subequations}
\noindent where $\porosity$ is the melt fraction, the volume fraction of liquid present within the two-phase material, and $\Gamma$ is the melting rate. We assume that the background melting timescale, which is in part controlled by the rate of mantle upwelling/downwelling, is much greater than the timescale of tidal deformation \cite{tackley2001ThreeDimensionalSimulations}, and henceforth take $\Gamma = 0$. For phase $j=\{l,s\}$, $\vec{v}_j$ is the velocity and $\rho_j$ is the density. Summing the above two equations gives the conservation of mass for the phase aggregate in terms of the phase-averaged density, $\rho = (1-\porosity)\rho_s + \porosity \rho_l$.

Neglecting inertial terms, the force balance equations for the solid and liquid phases are \cite{rovira-navarro2022TidesEnceladus}
\begin{subequations}\label{eq:gov_mom}
	\begin{align}
		\Div \brackets{\porosity \bm{\sigma}_l} - \porosity \rho_l \Grad \pot - \porosity \frac{\eta_l}{k} \vec{q} + p_l \Grad{\porosity} &= \vec{0} \label{eq:gov_mom_l} \\
		\Div \brackets{\brackets{1 - \porosity} \bm{\sigma}_s} - \brackets{1 - \porosity} \rho_s \Grad \pot + \porosity \frac{\eta_l}{k} \vec{q} - p_l \Grad{\porosity} \label{eq:gov_mom_s} &= \vec{0}
	\end{align}
\end{subequations}
\noindent where $\stressT_l$ and $\stressT_s$ are the liquid and solid stress tensors, respectively, $\Phi$ is the sum of all gravitational potentials, $k$ is the permeability of the solid skeleton, $\eta_l$ is the dynamic viscosity of the melt, and $\vec{q} \equiv \porosity (\vec{v}_l - \vec{v}_s)$ is the segregation flux. The second term in each equation represents the forces per unit volume of self- and external-gravitation, and the sum of the last two terms are known as the interphase force, $\vec{F} \equiv (\porosity \eta_l / k) \vec{q} - p_l \Grad{\porosity}$, the equal-and-opposite force per unit volume imposed by the melt on the solid phase \cite{katz2022DynamicsPartially}. The first term, the divergence of the solid and liquid stresses, balances the gravitation and interphase forces. Summing Eqs.~\ref{eq:gov_mom_l} and \ref{eq:gov_mom_s} cancels the interphase force and yields the force balance for the bulk mixture,
\begin{equation} \label{eq:gov_mom_bulk}
	\Div{\stressT} - \rho\Grad{\Phi} = 0,
\end{equation}
where the total stress is given by the volume-averaged sum of the solid and liquid stress tensors,
\begin{equation}\label{eq:total_stress_def}
	\stressT = (1-\porosity) \stressT_s + \porosity \stressT_l.
\end{equation}
For shorthand, we will sometimes denote the mean total stress as $\sigma \equiv \trace{\stressT}/3 = -p$, where $p$ is the total pressure. The permeable solid skeleton is assumed to be isotropically tortuous such that stresses in the fluid are also isotropic, 
\begin{equation} \label{eq:stress_liquid_def}
	\stressT_l = - p_l \identity,
\end{equation}
\noindent where $p_l \equiv -\trace{\stressT_l}/3$ is the pore pressure and $\identity$ is the identity matrix. The pore pressure is experienced by both the fluid and the solid grains as they are immersed in melt. 

Inserting Eq.~\ref{eq:stress_liquid_def} into \ref{eq:gov_mom_l} converts the melt's force balance into a modified form of Darcy's law,
\begin{equation}\label{eq:gov_darcy}
	\vec{q} = -M_\porosity \left( \Grad{p_l} + \rho_l \Grad{\pot} \right),
\end{equation}
where we have introduced the fluid's mobility, the ratio of permeability to liquid viscosity,
\begin{equation}\label{eq:mobility}
	M_\phi = k(\phi)/\eta_l.
\end{equation}
\noindent For a given pressure and gravitational forcing, a higher mobility results in a higher segregation flux. Mobility thus quantifies the liquid viscous resistance to melt circulation. We assume that $\eta_l = \SI{1}{\pascal\second}$, giving mobility and permeability a 1:1 correspondence throughout the manuscript. The mobility is a function of porosity due to the inherent dependence of permeability on the pore-fraction of the mantle, which we will examine further in Section \ref{ssec:total_heating_vs_melt_fraction}.

The strain of the material is defined as
\begin{equation}\label{eq:def_strain}
	\strainT_j \equiv \frac{1}{2}\left[\Grad{\dis_j} + \left(\Grad{\dis_j}\right)^T \right],
\end{equation}
\noindent where $\dis_j$ is the displacement of the solid ($j=s$) or liquid ($j=l$) relative to the pre-stressed, undeformed state. The deformation (strain) of the material must  now be linked to the stress it is subjected to via a constitutive law, which describes the stress response of the two-phase material to viscous and elastic strains. 

In the poro-viscous limit, the resistance to viscous (de)compaction is governed by the compaction viscosity, $\zeta$. This viscosity depends on both the melt fraction and shear viscosity of the two-phase aggregate (see~\S\ref{ssec:total_heating_vs_melt_fraction}). In the poroelastic limit, the resistance to elastic compaction is governed by the compaction modulus, $\kappa_d$, which is sometimes known as the drained bulk modulus and also depends on melt fraction. The compaction modulus is distinct from the bulk modulus of the solid grains, $\kappa_s$, because a two-phase material with incompressible solid grains can nonetheless compact elastically---by expelling or imbibing melt. The ratio of the compaction and bulk moduli is quantified by Biot's coefficient, $\alpha$, a constant that describes the compliance of the solid skeleton relative to the solid grains under isotropic stresses. Biot's coefficient is defined as
\begin{equation}\label{eq:biot_coeff_def}
	\alpha \equiv 1 - \frac{\kappa_d}{\kappa_s},
\end{equation} 
\noindent which is relevant for only elastic compaction. When $\alpha = 0$, the solid skeleton is as incompressible as the solid grains themselves, meaning that the skeleton is strong and effective at elastically resisting isotropic stresses.  
If $\alpha = 1$, the solid grains are incompressible and, relative to this, the skeleton is weak against isotropic stresses. As described next, we incorporate both viscous and elastic compaction behaviour in our constitutive law.

Forming a mechanically consistent constitutive law in the context of a two-phase material is less clear than for a single-phase material. It is therefore important to carefully formulate the problem, which we do following the original arguments presented in Chapter VII of ref.~\cite{love1911ProblemsGeodynamics}. The key argument made by Love \cite{love1911ProblemsGeodynamics} is that, upon displacement, a material element carries with it (i.e., advects) its initial hydrostatic stress. This motivates defining a constitutive law using material derivatives of the stress. For a two-phase material, extra caution is required; the solid and liquid phases that occupy a volume element in the deformed state may originate from different positions, thus carrying with them different hydrostatic pressures (even if the liquid and solid densities are equal). In supplementary information \ref{app:adv_maxwell}, we apply Love's argument to both poro-viscous and poroelastic rheological laws to derive an appropriate poro-viscoelastic constitutive law that captures advection of both the solid and liquid hydrostatic pressure fields. Our constitutive law for a compacting poro-viscoelastic Maxwell material is \cite{yang2000NonlinearViscoelastic,kervazo2021SolidTides,rovira-navarro2022TidesEnceladus,kamata2023PoroviscoelasticGravitational}
\begin{align}\label{eq:gov_maxwell_compaction}
	(1 - \phi) &\left[\frac{D_s \stressT_s}{D t} + \frac{1}{3}\frac{\kappa_d}{\zeta} \int_t \frac{D_s \trace{\stressT_s}}{D t}dt \identity\right] + \left[(\alpha - \phi)\frac{D_l p_{l}}{Dt} + (1 - \phi)\frac{\kappa_d}{\zeta} \int_t \frac{D_l p_l}{D t}dt \right] \identity \notag \\
	&+ (1 - \phi)\frac{\mu}{\eta} \left[\stressT_{s} - \frac{1}{3} \trace{\stressT_s}\identity\right] = \lambda_d \trace{\dot{\strainT}_s} \identity + 2 \mu \dot{\strainT}_s,
\end{align}
\noindent where $t$ is time, $\mu$ is the elastic shear modulus,  $\lambda_d = \kappa_d - 2\mu/3$, and $\eta,\,\zeta$ are the shear and compaction viscosities, respectively, which control the anelastic delay in the response of the two-phase material to shear and isotropic stresses. 
The Lagrangian derivatives for material phase $j=\{s,\,l\}$ are defined as
\begin{align}
	\frac{D_j }{Dt} \equiv \frac{\partial }{\partial t} + \vec{v}_j \cdot \Grad{}, 
\end{align}
\noindent and their appearance in Eq.~\ref{eq:gov_maxwell_compaction} is a result of applying Love's argument. Eq.~\ref{eq:gov_maxwell_compaction} reduces to the constitutive law used in \RNporo and \Kporo when $\zeta \rightarrow \infty$ (no viscous compaction) and when advective terms in the Lagrangian derivatives are neglected. The constitutive law in Eq.~6 of ref.~\cite{kervazo2021SolidTides} is recovered when there is no melt ($\alpha = \phi = 0$) and advective terms are neglected. The benefit and mathematical consistency of formulating the constitutive law with Lagrangian derivatives will become clear in the next section when the equations are linearised.  

The law relating the strain of the fluid to the liquid stress tensor can be derived using the following equations of state \cite{gassman1951UberElastizitat, biot1956TheoryPropagation},
\begin{subequations}\label{eq:eos}
	\begin{align}
		\frac{d\rho_l}{\rho_l} &= \frac{ d p_l}{\kappa_l} , \label{eq:eos_liquid} \\
		\frac{d \rho_s}{\rho_s} &= \frac{1}{\kappa_s} \left[ \frac{1}{1-\phi} d p^d + d p_l \right], 
	\end{align}  
\end{subequations}
\noindent where $p^d \equiv -\sigma - p_l$ is the differential between the total pressure and the pore pressure, and $\kappa_l$ is the bulk modulus of the liquid. Summing the liquid and solid continuity equations  \eqref{eq:gov_mass} and using the above equations of state \eqref{eq:eos} yields the storage equation,
\begin{equation}\label{eq:gov_storage}
	\phi \left[\frac{1}{\kappa_l} - \frac{1}{\kappa_s}\right] \frac{D_l p_l}{D t} + \frac{1}{\kappa_l} \vec{q} \cdot \Grad{p_l} -  \frac{1}{\kappa_s} \frac{D_s \sigma}{D t} +  \Div{\vec{q}}  +  \Div{\vec{v}_s} = 0,
\end{equation}
\noindent which we give in terms of the mean total stress $\sigma$. If we neglect all advective terms and assume that the two-phase material behaves elastically under isotropic stresses ($\zeta \rightarrow \infty$), then taking the trace of  Eq.~\ref{eq:gov_maxwell_compaction}  gives $\dot{\sigma} = \kappa_d \trace{\dot{\strainT}_s} + \alpha \dot{p}_l$, and $\Div{\vec{v}_s} = \trace{\dot{\strainT}_s}$, so Eq.~\ref{eq:gov_storage} reduces to
\begin{equation}\label{eq:gov_storage_elastic}
	S \frac{\partial p_l}{\partial t} + \alpha \trace{\dot{\strainT}_s}  = - \Div{\vec{q}} \qquad \text{(elastic isotropic deformation, $\zeta \rightarrow \infty$)},
\end{equation}
where $S$ is the storativity,
\begin{equation}\label{eq:storativity_def}
	S \equiv \frac{\phi}{\kappa_l} + \brackets{\alpha - \phi}\frac{1}{\kappa_s},
\end{equation}
the inverse of which is referred to as Biot's modulus \cite{biot1957ElasticCoefficients, kamata2023PoroviscoelasticGravitational}.
Equation \ref{eq:gov_storage_elastic} states that the compression of the liquid and solid on the left-hand side is balanced by changes in the relative volume flux of liquid entering or leaving the reference volume. We can see that if the melt becomes increasingly immobile, $\vec{q} \to \vec{0}$, then the right-hand side goes to zero and the pore pressure is dependent only on compaction of the solid skeleton. Integrating this expression with respect to time, assuming porosity is constant, and rearranging yields Eq.~3 in \Kporo and Eq.~17 in \RNporo when written in terms of the variation of fluid content, $\porosity \Divb{\vec{u}_l - \vec{u}_s}$. We note that while \Kporo does consider radial porosity variation throughout their manuscript, these variations are neglected in their storage equation. Following traditional poroelasticity, both \RNporo and \Kporo do not include advective terms in their storage equations. However, we proceed without neglecting the advective terms in Eq.~\ref{eq:gov_storage}; they originate from the material derivatives within the continuity equations, and therefore are essential in accounting for self-gravity and advection of hydrostatic stresses \cite{love1911ProblemsGeodynamics}. 

As we are interested in the more general storage equation that includes viscous compaction (finite $\zeta$) and advection of hydrostatic stress, we cannot easily time-integrate Eq.~\ref{eq:gov_storage} because of the integral terms present in Eq.~\ref{eq:gov_maxwell_compaction}. We can only further simplify the storage equation in the Fourier domain (\S\ref{ssec:theory_linearise}). 

Finally, Poisson's equation for the gravitational potential $\pot$ depends on the density of the mixture,
\begin{equation}\label{eq:gov_poisson}
	\nabla^2 \Phi = 4 \pi G \rho,
\end{equation}
where $G$ is the universal gravitational constant.  This equation must be solved everywhere.

The equations presented here all reduce to the standard solid-body deformation equations when $\porosity=0$ and $\alpha=0$ \cite{segatz1988TidalDissipation, sabadini2016GlobalDynamics}. 
Moderate simplification of the problem can be made by assuming incompressibility. If the solid is incompressible ($\kappa_s \to \infty$), then $\alpha \to 1$ and $S \to \porosity/ \kappa_l$.  If only the liquid is incompressible ($\kappa_l \to \infty$), then $S \to (\alpha - \phi)/ \kappa_s $. Assuming both phases are incompressible, the storage equation \eqref{eq:gov_storage} expresses mass conservation due to only segregation, $\Divb{\porosity \vec{v}_l} =  - \Div{(1-\porosity)\vec{v}_s}$. Even if both phases are incompressible, the compaction modulus of the two-phase aggregate can remain finite, meaning that the skeleton can still undergo volume changes due to expulsion/imbibition of melt.

\subsection{Linearisation and Fourier Transform}\label{ssec:theory_linearise}

In this section we linearise the dynamics that govern equations \ref{eq:gov_mass_l}, \ref{eq:gov_mass_s}, \ref{eq:gov_mom_bulk},  \ref{eq:gov_darcy}, \ref{eq:gov_maxwell_compaction}, \ref{eq:gov_storage}, and \ref{eq:gov_poisson}. 
We denote the base state with subscript $0$ and the perturbed (deformed) state with subscript $1$. The base state is assumed to be at rest, in hydrostatic equilibrium, and spherically symmetric such that it is only dependent on the radial coordinate $r = \left\vert \vec{r} \right\vert$. As the base state has $\vec{v}_0 = \vec{0}$, we omit the 1 subscript for $\vec{v}_j$ and $\vec{u}_j$, though they remain first-order quantities. The perturbed quantities can vary in three-dimensional space and so are functions of the position vector. That is, for any unknown $f_j$ of material phase $j=\{s\,,l\}$,
\begin{align}
	f_j(\pos, t) &= f_{j,0}(r) + f_{j,1}(\pos, t). \label{eq:perturb_form}  
\end{align}
The initial position of the material within the volume element at position $\vec{r}$ is less clear in the multi-phase problem than in the single-phase one. The subtlety to recognise is that the initial position of the material within that element can only be recovered through the displacement of the \textit{corresponding material phase} $j$.
The phase-dependent initial position $\pos_{j,0}$ is thus recovered through
\begin{align}
	\pos = \pos_{j,0} + \dis_j. \label{eq:pos_vec_def}
\end{align}
This approach is a natural extension of the argument made by Love \cite{love1911ProblemsGeodynamics} to the two-phase problem, and is crucial in ensuring mechanical consistency in the problem.

In the following, we take particular care in linearising the constitutive laws (Eqs.~\ref{eq:gov_maxwell_compaction} and \ref{eq:gov_storage}) as we form these in a different manner to previous works. When needed, the linearised equations are converted into the Fourier domain by assuming time-periodic solutions of the form, 
\begin{equation}\label{eq:fourier_def}
	f_{j,1} = \f{f}_{j,1} e^{i \omega t},
\end{equation} 
where $\f{f}_j$ is the complex Fourier-transformed variable of material phase $j$, the forcing frequency is $\omega$, and $i$ is the imaginary number. 

Expanding the density field in the form of Eq.~\ref{eq:perturb_form} gives,
\begin{equation}
	\rho_j(\pos, t) = \rho_{j,0}(r) + \rho_{j,1} (\pos, t). \label{eq:perturb_den_decomp}
\end{equation}
Inserting this into the conservation of mass equations (\ref{eq:gov_mass}), neglecting second-order terms in perturbed quantities, and integrating with respect to time gives the density perturbations,
\begin{subequations}
	\begin{align}
		\phi \rho_{l,1}  &= - \disL \cdot \Grad \brackets{ \phi \rho_{l,0} }- \phi \rho_{l,0}  \Div{\bm{u}_l}, \label{eq:perturb_den_l} \\
		\brackets{1 - \porosity} \rho_{s,1} &= - \disS \cdot \Grad\left[\brackets{1 - \porosity} \rho_{s,0} \right] - \brackets{1 - \porosity} \rho_{s,0} \Div{\bm{u}_s},\label{eq:perturb_den_s}
	\end{align}\label{eq:perturb_den}
\end{subequations}
where we have assumed that porosity is constant (discussed in supporting information \ref{app:porosity_assumption}). The density perturbations thus consist of two parts: a perturbation due to advection of the base state, and a perturbation due to compression or expansion of the solid or liquid. The advection term is critical to account for alteration of the gravitational potential due to mass redistribution.

Expanding the gravitational potential in the form of Eq.~\ref{eq:perturb_form} gives,
\begin{equation}
	\pot(\pos, t) = \pot_0(r) + \pot_1(\pos, t), \label{eq:perturb_pot_decomp}
\end{equation}
\noindent where $\pot_0$ is the pre-stressed gravitational potential of the body. At rest and in the absence of tidal forcing, Poisson's equation \eqref{eq:gov_poisson} for the base state gravitational potential integrates to
\begin{equation}
	\nabla \pot_0(r) = g(r) \rhat, \label{eq:perturb_pot_0}
\end{equation}
\noindent where $\rhat$ is the radial basis vector and $g(r) = G \int_V \rho dV / r^2$, where $V$ is the spherical volume of radius $r$. Poisson's equation for the gravitational perturbation is found by inserting equations \ref{eq:perturb_pot_decomp} and \ref{eq:perturb_den} into \ref{eq:gov_poisson},
\begin{equation}
	\nabla^2 \pot_1 = -4\pi G \left[u_s^r \partial_r \den_0 + \den_{0}\Div{\disS} + u_{\text{rel}}^r \partial_r\brackets{\phi\denL{0}} + \phi\denL{0}\Div{\disR}\right], \label{eq:perturb_pot_1}
\end{equation}
where $\vec{u}_{\text{rel}} \equiv \dis_l - \dis_s$ is the relative/segregation displacement, superscript $r$ indicates a radial vector component, and $\partial_r$ is the partial derivative with respect to radius. This expression is equivalent to Eq.~21 in \Kporo, and Eq.~33 in \RNporo under the assumption of uniform layer density and melt fraction.

The solid Cauchy stress tensor and pore pressure are decomposed identically to Eq.~\ref{eq:perturb_form},
\begin{subequations}\label{eq:perturb_stress_decomp}
	\begin{align}
		\stressT_s(\pos, t) &= \stressT_{s,0}(r)  + \stressT_{s,1}(\pos, t),   \\
		p_l(\pos, t) &= p_{l,0}(r)  + p_{l,1}(\pos, t),  
	\end{align}
\end{subequations}
which consists of only the hydrostatic base state and a perturbation due to deformation. The decomposition in Equation \ref{eq:perturb_stress_decomp} is identical to the decomposition of $\Phi$ and $\rho_j$, which we emphasise is a different approach to most tidal deformation texts where an advective term is forced into the decomposition \cite{sabadini2016GlobalDynamics}. Below, we show that the approach used here reproduces the classic tidal deformation equations because material advection is naturally accounted for in our constitutive laws (Eqs.~\ref{eq:gov_maxwell_compaction} and \ref{eq:gov_storage}). 

At rest  ($p_l = p_{l,0}$, $\stressT_s = \stressT_{s,0}$, $\stressT = \stressT_0$) and using Eqs.~\ref{eq:stress_liquid_def} and \ref{eq:perturb_pot_0}, the liquid, solid, and bulk momentum equations (\ref{eq:gov_mom} and \ref{eq:gov_mom_bulk}) reduce to hydrostatic balance,
\begin{subequations} \label{eq:perturb_mom_0}
	\begin{align}
		-\Grad{p_{l,0}} &= \rho_{l,0} g \rhat, \label{eq:perturb_mom_l0} \\
		-\Grad{p_{s,0}} &= \rho_{s,0} g \rhat, \label{eq:perturb_mom_s0} \\
		\Div{\stressT_0} = - \Grad{p_0} &= \rho_0 g \rhat, \label{eq:perturb_mom_b0}
	\end{align}
\end{subequations}
where in Equation \ref{eq:perturb_mom_s0} we have assumed that porosity is a constant.

Inserting Eq.~\ref{eq:perturb_stress_decomp} into the solid constitutive law (Eq.~\ref{eq:gov_maxwell_compaction}), neglecting second-order terms in perturbed quantities, and assuming hydrostatic balance in the base state by inserting Eq.~\ref{eq:perturb_mom_0} gives,
\begin{align}\label{eq:perturb_const_s_2}
	\dot{\stressT}_{1} &+ \alpha \dot{p}_{l,1} \identity + \left[(\rho_{0} - \alpha\rho_{l,0}) v_s^r  - (\alpha - \phi)  v_{rel}^r \rho_{l,0}   \right]g\identity + \frac{\mu}{\eta} \left[\stressT_{1} - \frac{1}{3} \trace{\stressT_{1}}\identity\right] \notag \\ 
	&+ \frac{\kappa_d}{\zeta} \left[\frac{1}{3} \trace{\stressT_{1}} + p_{l,1} + (\rho_{0} - \phi\rho_{l,0}) u_s^r g  - (1 - \phi)  u_{rel}^r \rho_{l,0} g\right]\identity = \lambda_d \trace{\dot{\strainT}_s} \identity + 2 \mu \dot{\strainT}_s,
\end{align}
\noindent which is written in terms of the total stress perturbation and solid and relative velocities and displacements. Here, overdots represent partial time derivatives, $\partial/\partial t$. This equation is more compact in the frequency domain; taking the Fourier transform of Eq.~\ref{eq:perturb_const_s_2} and rearranging gives
\begin{equation}\label{eq:fourier_maxwell_compaction}
	\f{\stressT}_1   = \left[ \f{\kappa}_d  - \frac{2}{3}\f{\mu} \right]  \trace{\f{\strainT}_s} \identity + 2\f{\mu} \f{\strainT}_s - \f{\alpha}\f{p}_{l,1}\identity - \left[(\rho_0-\f{\alpha}\rho_{l,0})\f{u}_s^r  - (\f{\alpha} - \phi)\f{u}_{rel}^r \rho_{l,0}\right]g\identity,
\end{equation}
where the complex shear modulus, drained compaction modulus, and Biot's coefficient are,
\begin{subequations}
	\begin{align}
		\f{\mu} &\equiv \frac{i \omega \mu}{i \omega + \mu/\eta}, \\
		\f{\kappa}_d &\equiv \frac{i \omega \kappa_d}{i \omega + \kappa_d/\zeta}, \\
		\f{\alpha} &\equiv 1 - \frac{\f{\kappa}_d}{\kappa_s},
	\end{align}
\end{subequations}
respectively. The mean total stress perturbation is given by the trace of Equation \ref{eq:fourier_maxwell_compaction},
\begin{subequations}
	\begin{align}
		\trace{\f{\stressT}_1}/3 = \f{\sigma}_1 &= \f{\kappa}_d \trace{\f{\strainT}_s} - \f{\alpha} \f{p}_{l,1} - \left[(\rho_0 - \f{\alpha}\rho_{l,0})\f{u}_s^r  - (\f{\alpha} - \phi)\f{u}_{rel}^r \rho_{l,0}\right]g. \label{eq:fourier_maxwell_trace_total}
	\end{align}
\end{subequations}
Next, we linearise the storage equation in \ref{eq:gov_storage} and insert Eqs.~\ref{eq:perturb_mom_l0} and \ref{eq:perturb_mom_b0}, 
\begin{align}\label{eq:perturb_storage} 
	\left[\frac{\phi}{\kappa_l} - \frac{\phi}{\kappa_s} \right] \left[ \dot{p}_{l,1} - v_l^r \rho_{l,0} g \right] &=  \frac{\phi}{\kappa_s} v^r_{rel} \rho_{l,0} g  +  \frac{1}{\kappa_s} \left[\dot{\sigma}_1 + \rho_0 v_s^r g \right] -  \phi \trace{\dot{\strainT}_{rel}}  -  \trace{\dot{\strainT}_{s}} , 
\end{align}
\noindent where $\dot{\sigma}_1$ is obtained by taking one third the trace of Equation \ref{eq:perturb_const_s_2} (see supporting information \ref{app:storage}). 
Assuming Fourier solutions of the form in \ref{eq:fourier_def} for Equation \ref{eq:perturb_storage}, inserting Eq.~\ref{eq:fourier_maxwell_trace_total} and rearranging gives
\begin{equation}\label{eq:fourier_storage}
	\f{p}_{l,1} = \f{u}_l^r \rho_{l,0} g  - \f{S}^{-1} \left[ \f{\alpha}   \trace{\f{\strainT}_s} +  \phi\trace{\f{\strainT}_{rel}}\right],
\end{equation}
where the complex storativity is,
\begin{equation}\label{eq:fourier_storativity_def}
	\f{S} \equiv \frac{\phi}{\kappa_l} + \brackets{\f{\alpha} - \phi}\frac{1}{\kappa_s}.
\end{equation}
\noindent For later convenience and easier comparison to existing work, we write the pore pressure perturbation as
\begin{subequations}\label{eq:fourier_pore_pres}
	\begin{equation}\label{eq:fourier_pore_pres_adv}
		\f{p}_{l,1} = \f{u}_l^r \rho_{l,0} g  + \f{p}_{l,1}^\delta,
	\end{equation} 
	where the non-advective portion of the liquid stress is defined as
	\begin{equation}\label{eq:fourier_pore_pres_nonadv}
		\f{p}_{l,1}^\delta \equiv -\f{S}^{-1} \left[   \f{\alpha}   \trace{\f{\strainT}_s} +  \phi\trace{\f{\strainT}_{rel}}\right].
	\end{equation}
\end{subequations}
Equation \ref{eq:fourier_pore_pres_nonadv} is equivalent to Equation 6 in \Kporo if compaction is ignored ($\zeta \rightarrow \infty$). We can then insert Eq.~\ref{eq:fourier_pore_pres_adv} into the linearised frequency-domain solid constitutive law, Eq.~\ref{eq:fourier_maxwell_compaction}, which simplifies to
\begin{subequations}\label{eq:fourier_maxwell_compaction2}
	\begin{equation}\label{eq:fourier_maxwell_compaction2_adv}
		\f{\stressT}_1   =  - \left[\rho_0\f{u}_s^r  +  \phi\f{u}_{rel}^r \rho_{l,0}\right]g\identity + \f{\stressT}_1^\delta,
	\end{equation}
	\noindent where we have defined the non-advective stress perturbation,
	\begin{equation}\label{eq:fourier_maxwell_compaction2_nonadv}
		\f{\stressT}_1^\delta   \equiv \left[ \f{\kappa}_d  - \frac{2}{3}\f{\mu} \right]  \trace{\f{\strainT}_s} \identity + 2\f{\mu} \f{\strainT}_s - \f{\alpha}\f{p}^\delta_{l,1}\identity.
	\end{equation}
\end{subequations}
Equation \ref{eq:fourier_maxwell_compaction2_adv} is the frequency-domain poro-viscoelastic constitutive law that includes advection of the hydrostatic base state pressure field, and is equivalent to Eq.~41 in \Kporo. As expected from the correspondence principle, Equations \ref{eq:fourier_maxwell_compaction2_adv} and \ref{eq:fourier_maxwell_compaction2_nonadv} are analogous to the poroelastic problem in the time domain. If there is no melt, $\phi=0$, $\kappa_d = \kappa_s$, and therefore $\alpha=0$, Equation \ref{eq:fourier_maxwell_compaction2} reduces to the elastic constitutive law in Eq.~6, Chapter VII of ref.~\cite{love1911ProblemsGeodynamics}.

Now that we have a frequency-domain constitutive law for the liquid and solid (Eqs.~\ref{eq:fourier_pore_pres_adv}, \ref{eq:fourier_pore_pres}, \ref{eq:fourier_maxwell_compaction2}), we can proceed to linearise and Fourier transform the momentum equations.

The linearised frequency-domain liquid momentum equation is obtained by substituting Eqs.~\ref{eq:stress_liquid_def}, \ref{eq:perturb_den_decomp}, \ref{eq:perturb_den}, \ref{eq:perturb_pot_decomp}, \ref{eq:perturb_pot_1}, \ref{eq:perturb_stress_decomp}, \ref{eq:perturb_mom_l0}, and \ref{eq:fourier_pore_pres_adv} into the Fourier transform of Darcy's law in Eq.~\ref{eq:gov_darcy}. Neglecting second order terms in perturbed quantities, this gives
\begin{align}
	i \omega \porosity \f{\vec{u}}_{rel} &= -M_\porosity \left[\rho_{l,0} \partial_r\left( g \f{u}_{l}^r\right)\rhat + \Grad{\f{p}_{l,1}^\delta}   +\rho_{l,0} \Grad{\f{\Phi}_1}  - \rho_{l,0} g\left(\Div{\f{\vec{u}}_l}\right)\rhat\right], \label{eq:perturb_mom_l1}
\end{align}
\noindent which is a form of Darcy's law that accounts for compression and self-gravity of the circulating melt. Equation \ref{eq:perturb_mom_l1} is identical to Eq.~49 in \Kporo when ignoring inertial terms, radial variations in porosity,  and setting $u_l^r = u^r_{\text{rel}} + u_s^r$ and $\Div{\f{\dis}_l} = \trace{\f{\strainT}_s + \f{\strainT}_{rel}}$. As noted in \Kporo, this differs from the Fourier transform of Eq.~29b in \RNporo because advection of the base state pressure and density fields was neglected. We show in supporting information \ref{app:checks} that this leads to slightly incorrect energy dissipation rates.

Following the same procedure for the Fourier transform of the momentum equation of the bulk mixture (Eq.~\ref{eq:gov_mom_bulk}) gives
\begin{align}
	\Div{\f{\stressT}_1^\delta} &= \rho_{0} \partial_r\brackets{ g \f{u}^r_{s} }\rhat + \phi \rho_{l,0} \partial_r \brackets{g \f{u}^r_{rel}} \rhat + \rho_0\Grad{\f{\Phi}_1}  - \left[\rho_{0}\Div{\f{\vec{u}}_s}  + \phi \rho_{l,0}\Div{\f{\dis}_{rel}}\right]g \rhat. \label{eq:perturb_mom_s1}
\end{align}
\noindent This equation is identical to Eq.~48 in \Kporo if inertial terms are neglected. Moreover, our Equation \ref{eq:perturb_mom_s1} differs from the Fourier transform of \RNporo Eq.~29a because advection of the base state pressure and density fields was neglected. In their case, Eq.~29a in \RNporo does not reduce to the classic, solid-body, linearised momentum equation if $\porosity = \rho_l = 0$ (Eq.~1.58 in ref.~\cite{sabadini2016GlobalDynamics}). This is because of the assumption of uniform layer density in \RNporo.

Despite having formulated the solid and liquid constitutive laws to explicitly include material advection---a different approach to most tidal deformation studies---we have arrived at the same set of linearised momentum equations. We argue that this approach, which is simply an extension of the classic elastic approach by Love \cite{love1911ProblemsGeodynamics} to the viscoelastic and poro-viscoelastic problems, is more physically intuitive and mathematically consistent than that outlined in, for example, ref.~\cite{sabadini2016GlobalDynamics}. This consistency is the inclusion of material advection in our constitutive laws, which avoids the need to make an ad-hoc assumption about why stresses should be linearised differently to other problem variables, as is commonly done.

The governing equations have now been linearised and converted to the frequency domain (Equations \ref{eq:perturb_pot_1}, \ref{eq:fourier_pore_pres}, \ref{eq:fourier_maxwell_compaction2}, \ref{eq:perturb_mom_l1} and \ref{eq:perturb_mom_s1}). To solve these coupled equations, the unknown quantities are expanded in spherical harmonics, converted to a set of ODEs in radius, and then numerically integrated. Details regarding this are well documented in the literature; we provide further information in the supporting information \ref{app:ODE} and \ref{app:solve_ODEs}.

\subsection{Dissipation}\label{ssec:theory_dissipation}

After the tidal response of Io is computed, we determine the tidal dissipation rate throughout the body. The dissipation rate is inherently related to the time-delay in the response of the body to the forcing potential, and is thus dependent on the imaginary parts of the complex rheological parameters, $\f{\mu}$, $\f{\kappa}_d$, $\f{S}$, as  well as $M_\porosity$.

The forcing period-averaged dissipation rate per unit volume is \cite{liao2020HeatProduction},
\begin{equation}\label{eq:diss_def}
	\dot{E}^v = \frac{1}{P} \int_P \underbrace{\stressT_1 : \dot{\strainT}_s - \phi p_l \trace{\dot{\strainT}_{rel}}}_{solid} + \underbrace{\frac{\vec{q}\cdot\vec{q}}{M}}_{liquid} dt,
\end{equation} 
where Io's 42-hour forcing period is $P = 2\pi /\omega$. The first two terms on the right-hand represent dissipation due to shear and compaction of the solid skeleton itself. The last term accounts for viscous heating in the melt, otherwise known as Darcy dissipation. We refer to heating from isotropic deformation as \textit{compaction} dissipation, rather than the commonly used bulk dissipation \cite{kervazo2021SolidTides}. This choice avoids the notion that heating from isotropic deformation is a phase- or spatially averaged quantity, such as the bulk density. 
Similar to \Kporo, we can split this dissipation rate into contributions from shearing $\dot{E}^v_S$, compaction $\dot{E}^v_{C}$, and (viscous) Darcy flow due to fluid--solid segregation $\dot{E}^v_{D}$;
\begin{subequations}\label{eq:diss_all}
	\begin{align}
		\dot{E}^v_S &= -\Omega \Im{\f{\mu}} \left[
		\left|\f{\epsilon}_{rr}\right|^2+ \left|\f{\epsilon}_{\theta\theta}\right|^2+\left|\f{\epsilon}_{\varphi\varphi}\right|^2
		+ 2\left|\f{\epsilon}_{r\theta}\right|^2 + 2\left|\f{\epsilon}_{r\varphi}\right|^2 + 2\left|\f{\epsilon}_{\theta\varphi}\right|^2
		-
		\frac{1}{3}\left|\trace{\strainT_s}\right|^2 
		\right], \label{eq:diss_sol_shear}\\ 
		\dot{E}^v_{C} &= -\frac{\Omega}{2}\Im{\f{\kappa}_d} \left[ \left|\trace{\f{\strainT}_s}\right|^2 + \left|\frac{\f{p}_l}{\kappa_s} \right|^2\right], \label{eq:diss_bulk} \\
		\dot{E}^v_{D} &=  \frac{\phi^2\Omega^2}{2}\frac{1}{M} \left|\f{\vec{u}}_{rel}\right|^2 \label{eq:diss_liq_mobility},
	\end{align}
\end{subequations} 
where $\Omega = | \omega |$ is Io's rotation frequency. In contrast to \Kporo, we have combined fluid and solid effects in our expression for compaction dissipation \eqref{eq:diss_bulk}. While Eq.~\ref{eq:diss_bulk} contains contributions from both solid deformation and melt, the resulting dissipation only occurs in the solid. 
Compaction dissipation therefore represents the heating associated with solid-grain rearrangement as pore space is opened/closed.   
The total time-averaged dissipation rate is the sum of Eqs.~\ref{eq:diss_all} over the total volume of the body,
\begin{equation}\label{eq:diss_E_integrate}
	\dot{E} = \int_V \left( \dot{E}^v_S + \dot{E}^v_{C} + \dot{E}^v_{D} \right)dV. 
\end{equation}

The strain, displacement and pressure magnitudes in Eq.~\ref{eq:diss_all} depend on the type, frequency, and magnitude of the tidal forcing. We force Io with the time-varying gravitational potential arising from small orbital eccentricity, $e$, limited to the diurnal (orbital) frequency \cite{kaula1961AnalysisGravitational}. We neglect the much smaller forcing due to obliquity and adjacent moons \cite{hay2020PoweringGalilean, hay2022HighFrequency}. Specific details are given in supporting information \ref{app:forcing}.

The total dissipation rate can also be determined using the phase-delay of the gravitational response of the body, given by the imaginary component of the tidal Love number $k_2$. This can be recovered with the response gravitational potential computed at the surface (see supplementary information Eq.~\ref{eq:k2}). For a synchronously rotating body in an eccentric orbit with eccentricity $e \ll 1$, the total dissipation rate is \cite{segatz1988TidalDissipation}
\begin{equation}\label{eq:diss_E_k2}
	\dot{E} = - \frac{21}{2} \Im{k_2} \frac{\Omega^5 R^5}{G} e^2.
\end{equation} 
A crucial check to the numerical solution of the tidal deformation problem is that Equations \ref{eq:diss_E_integrate} and \ref{eq:diss_E_k2} give identical results. 

Now that the mathematical problem has been posed, we must next define the internal and rheological structure of Io.

\section{Internal Structure and Boundary Conditions} \label{sec:internal_structure}

\begin{table}[t]
	\small
	\centering
	\begin{tabular}{lllcccc}
		\hline 
		Quantity  & Symbol & Units & Crust & Asthenosphere & Lower Mantle & Core \\ \hline
		Layer thickness & $\Delta r$ &\si{\km} & 20 & 50--400 & 700--1050 & 700 \\
		Solid density &$\rho_s$ &\si{\kg\per\metre\cubed} & 3000 & 3300 & 3300 & - \\
		Liquid density &$\rho_l$ &\si{\kg\per\metre\cubed} & - & 3300 & - & 7640 \\
		Shear modulus &$\mu$ &\si{\giga\pascal} & 60 & $60^*$ & $60$ & 0 \\
		Solid bulk modulus &$\kappa_s$ &\si{\giga\pascal} & 200 & 200 & 200 & - \\
		Liquid bulk modulus &$\kappa_l$ &\si{\giga\pascal} & - & 1--200 & - & 200 \\
		Shear viscosity &$\eta$ & \si{\pascal\second} & \num{e25} & \num{e12}--\num{e21}$^*$ & \num{e21} & - \\
		Liquid viscosity &$\eta_l$ & \si{\pascal\second} & - & 1 & - & 0 \\
		Biot's coefficient &$\alpha$ & - & - & 0.01--1$^*$ & - & - \\
		Compaction modulus & $\kappa_d$ & \si{\giga\pascal} & - & 2--200$^*$ & - & -  \\
		Permeability & $k$ & \si{\square\metre} & - & \num{e-10}--\num{e-3}$^*$ & - & - \\\hline
	\end{tabular}
	\caption{Parameters describing Io's internal structure and corresponding material properties.  The compaction modulus $\kappa_d$ is controlled by $\alpha$ through Equation \ref{eq:biot_coeff_def} or \ref{eq:biot_coeff_phi}. Asterisks indicate parameters that can vary with melt fraction.}
	\label{tb:io_interior}
\end{table}

Io's internal structure has been constrained through determination of its gravity field, moment of inertia, topography, and tidal deformation by the Galileo and Juno spacecraft \cite{anderson1996GalileoGravity,anderson2001IoGravity, park2024IosTidal}. Its inferred moment of inertia is consistent with a differentiated body with metallic iron core \cite{anderson2001IoGravity}. The tidal perturbation of its gravity field is consistent with an interior where the melt fraction is below the disaggregation limit (i.e., $\phi < 0.3$), meaning no shallow magma ocean is present \cite{park2024IosTidal, aygun2025LoveNumbers}. Mountains with relief of \SI{20}{\km} indicate that the lithosphere is strong, despite Io's abundant surface volcanism \cite{schenk1998OriginMountains, turtle2001MountainsIo}. To respect these constraints while retaining simplicity, we adopt a four-layer internal structure.  This structure, quantified in Table \ref{tb:io_interior}, comprises a molten core, solid lower mantle, partially molten asthenosphere, and a solid, stiff lithosphere. 
We assume that the origin of the asthenosphere's melt is localised tidal heating, thus we consider only a ``shallow'' mantle model of dissipation, neglecting contributions from the deep mantle \cite{segatz1988TidalDissipation}.
We vary the thickness $H$ and melt fraction $\porosity$ of the asthenosphere, as well as the mobility of the melt--solid aggregate (Eq.~\ref{eq:mobility}) and compressibility of the melt. The solid and melt densities are taken to be equal ($\rho_s = \rho_l$), ensuring that the base state of the asthenosphere is at rest, simplifying our analysis. 

We know from observation and modelling that melt must pass from asthenosphere into the lithosphere where it is erupted onto the surface \cite{davies2003VolcanismIo,moore2001ThermalState}. However, over the timescale of diurnal tides, flow in or out of the asthenosphere is likely small. We therefore take the top and bottom boundaries of the asthenosphere to be impenetrable such that $\vec{q}\cdot \vec{e}_r = 0$. Melt is thus confined to the asthenosphere in our model. For other possible boundary conditions, see \RNporo and \Kporo.

\section{Results} \label{sec:results}

In Section \ref{sec:theory}, we presented the two-phase theory of gravito-poroviscoelastic dynamics. In this section, we apply the theory to Io, assuming an internal structure as given in \S\ref{sec:internal_structure}. To begin, 
in \S\ref{ssec:overview} we gain intuition for the primary controls and broad patterns of Io's two-phase tidal response. For this, we use a simplified reference case with incompressible solid and liquid phases. In \S\ref{ssec:heating}, we relax the incompressibility assumption and determine the sensitivity of the tidal heating solutions to other two-phase control parameters. We then allow the poro-viscoelastic parameters to co-vary as functions of $\porosity$ in \S\ref{ssec:total_heating_vs_melt_fraction}.  We treat the melt fraction as an input parameter, and assume that it is constant and uniform. Calculations with different values of $\porosity$ enable a more realistic insight into how Io's heat generation might be influenced by its melt fraction. \S\ref{ssec:patterns} presents the spatial patterns of heating associated with each heating mechanism, and how these compare to the distribution of Io's volcanoes. The code to reproduce all figures is available at \cite{hay2025JuliaCode}.

\subsection{Primary controls on tidal deformation and heating}\label{ssec:overview}

\begin{figure}[t]
	\centering
	\includegraphics[width=\linewidth]{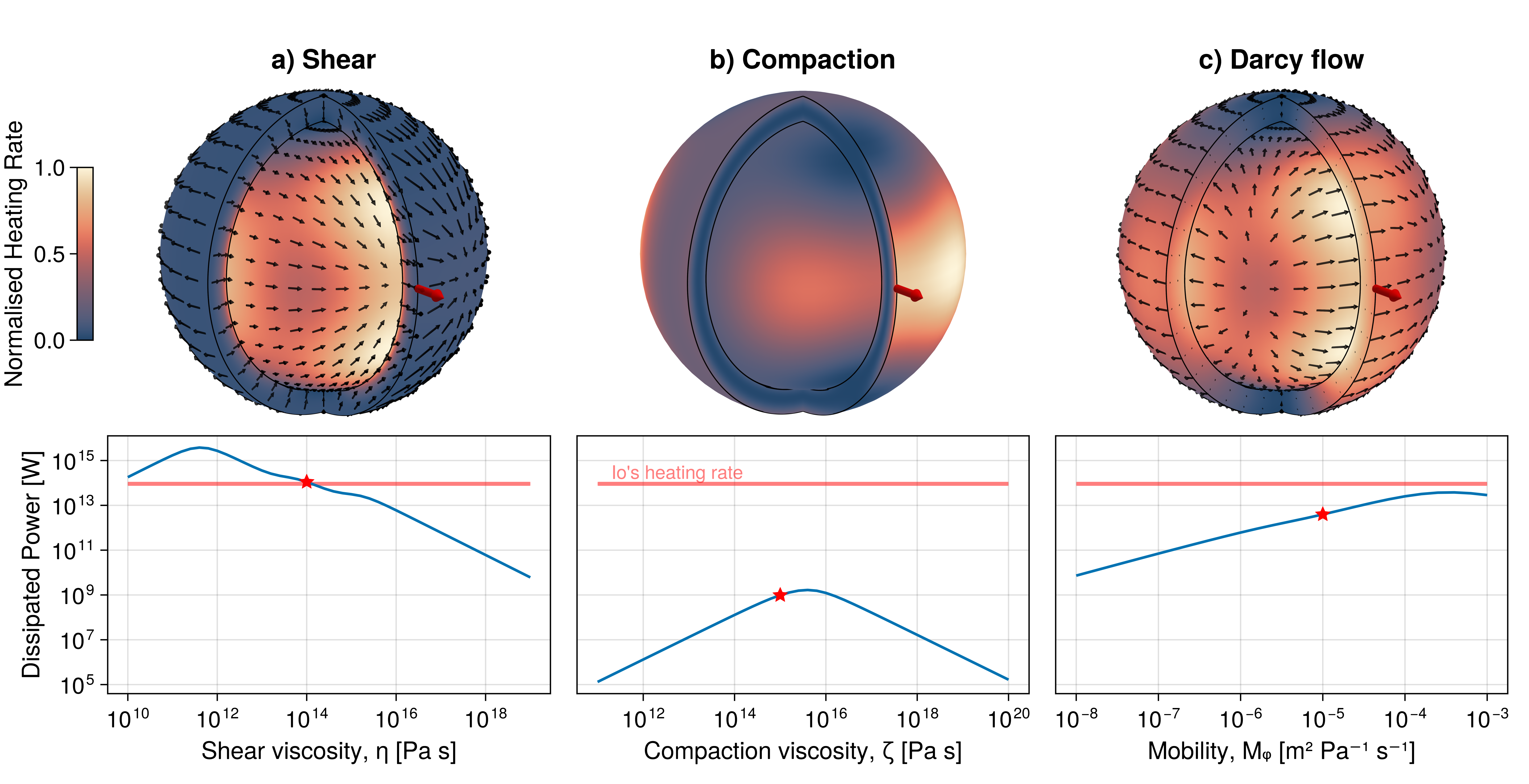}
	\caption{Qualitative tidal deformation solutions of a 300 km thick asthenosphere at perijove ($t=0$) for deformation due to \textbf{a)} shear, \textbf{b)} compaction, and \textbf{c)} melt segregation. The arrows in \textbf{a)} and \textbf{c)} indicate the solid and segregation displacements, respectively ($\vec{u}_s$ and $\vec{u}_{rel}$), tangent to the surface in which they are plotted. The tidal axis (red arrow) points towards Jupiter. The inner and outer sphere are the base and surface of the asthenosphere, respectively, and the colours represent the period-averaged volumetric heating rate within each surface plotted. Panels \textbf{d)}, \textbf{e)} and \textbf{f)} show the total heating rate for shear, compaction, and Darcy flow, respectively, as a function of each mechanism's primary control parameter. Red stars correspond to the solution shown in the top row. Io's observed heating rate (red line) is from ref.~\cite{lainey2009StrongTidal}.}
	\label{fig:overview}
\end{figure}

A set of reference, two-phase tidal deformation solutions within an $H =$~300~km thick asthenosphere are shown in Figure \ref{fig:overview}. Here, we assume that each phase is incompressible, $\kappa_l \to \infty$, $\kappa_s \to \infty$, while the two-phase aggregate remains compactible, $\kappa_d = 200$~GPa, so that $\alpha \to 1$.

In Figure \ref{fig:overview}a we see the tidal deformation solutions associated with shear, which is the largest part of Io's tidal deformation. The total heating rate varies with shear viscosity (bottom panel), with a peak close to where the shear Maxwell time $\tau_S = \eta / \mu$ of the asthenosphere approaches the forcing period, typical of many tidal heating studies \cite{segatz1988TidalDissipation}. The top panel shows the asthenosphere's displacement field with a shear viscosity of $\eta=\num{e14}$~Pa~s, which reproduces Io's total heating rate (star, bottom panel). This shear viscosity is much lower than that of Earth's mantle, which we discuss further in \S\ref{sec:discussion}. We see a clear angular separation between the tidal axis (red arrow), and the region of highest tide (area of convergence at the surface). The volumetric tidal heating rate peaks at the base of the asthenosphere, and has a dissipation pattern with maxima at mid-latitudes and zero at the poles, consistent with the ``shallow mantle'' model of ref.~\cite{segatz1988TidalDissipation}.

Heating due to compaction is shown in Figure \ref{fig:overview}b. Like shear deformation, the total compaction-heating rate peaks at a specific value of $\zeta$. This peak occurs when the \textit{compaction} Maxwell time $\tau_C = \zeta / \kappa_d$ approaches the forcing period. When $\zeta$ is high, compaction is dominantly elastic, and when $\zeta$ is low, compaction is viscous. The peak heating rate occurs between these two limits. However, for any $\zeta$, heating from compaction is substantially below Io's observed heating rate. This is a consequence of assuming a fairly incompactible asthenosphere ($\kappa_d = 200$~GPa) and incompressible solid and liquid phases, as will be shown in \S\ref{ssec:heating}\ref{sssec:heating_bulk}. The time-averaged heating pattern, consistent with ref.~\cite{kervazo2021SolidTides}, is focused towards the equator, peaks either side of the tidal axis, and is non-zero at the poles. For this reference case, heating is focused in the shallow asthenosphere, unlike the distribution of shear heating.

Finally, in Figure \ref{fig:overview}c, we see the liquid displacement and heating rates due to Darcy flow (solid--melt segregation). 
The Darcy displacement at $t=0$ for $M_\phi = \SI{e-5}{\metre\squared\per\pascal\per\second}$ is shown in the top row. We see that the melt primarily segregates laterally, flowing (relative to the solid) from one side of the tidal axis to the other. In this reference model, the lateral volumetric heating pattern is similar to that from shear deformation, but is much more uniform over the depth of the asthenosphere. 
The total Darcy heating rate, shown in the bottom panel, is controlled by the mobility of the melt, $M_\phi$. Similarly to shear and compaction, there is a critical mobility at which the Darcy heating rate peaks, with decreasing dissipation to either side of this value.
The critical mobility results from a competition between the forces that oppose elastic compaction: viscous stresses in the melt, and elastic resistance of the solid skeleton. 
When all phases are incompressible, compaction can only be accommodated through melt segregation. This is shown by the storage equation (\ref{eq:gov_storage}), which reduces to $\Div{\vec{q}} = - \Div{\vec{v}_s}$ when $\kappa_l \to \infty$ and $\kappa_s \to \infty$.
When the mobility is small, viscous resistance in the melt is high, preventing melt from segregating quickly, and consequently the compaction rate is low. In this regime, $|\vec{q}| \propto M_\phi$ and Darcy dissipation increases with mobility (Eq.~\ref{eq:diss_liq_mobility}). 
As the mobility further increases, the melt's viscous stresses that resist segregation and compaction decrease. When the mobility is sufficiently large, compaction is no longer limited by viscous resistance to melt segregation, but instead by elastic resistance of the skeleton to isotropic stresses, controlled by the elastic compaction modulus $\kappa_d$. 
The critical mobility occurs when this elastic resistance prevents any further compaction of the solid. 
At this point, mass conservation forces the segregation flux to become constant (enforced by an increase in pore pressure gradient), while the heat producing viscous resistance to segregation decreases with $1/M_\porosity$ (Eq.~\ref{eq:diss_liq_mobility}). 
Consequently, Darcy dissipation decreases with $M_\porosity$ past this point.

\subsection{Influence of compressibility on tidal heating rates}
\label{ssec:heating}

In this section, we allow the solid and liquid phases to be compressible, while still imposing a constant melt fraction. We investigate the sensitivity of our tidal heating calculations to the bulk modulus of the liquid $\kappa_l$, as well as the compressibility of the two-phase aggregate (relative to the solid grains), by varying Biot's modulus $\alpha$. We will refer to a ``strong'' asthenosphere as one with small $\alpha$, and a ``weak'' asthenosphere when $\alpha$ is large. Strong or weak in this context reflects the elastic resistance of the asthenosphere to compaction (isotropic) stresses, relative to a purely solid asthenosphere.  The bulk modulus of the solid grains is held constant at $\kappa_s = 200$~GPa. 

We first determine the tidal dissipation sensitivity to compressible effects for shear heating in the poro-viscoelastic limit (\S\ref{sssec:heating_shear}). We then consider compaction heating while enforcing elastic shear (\S\ref{sssec:heating_bulk}). Finally, we investigate Darcy heating in the poroelastic limit (\S\ref{sssec:heating_darcy}).

\begin{figure}[!t]
	\centering
	\includegraphics[width=0.85\linewidth]{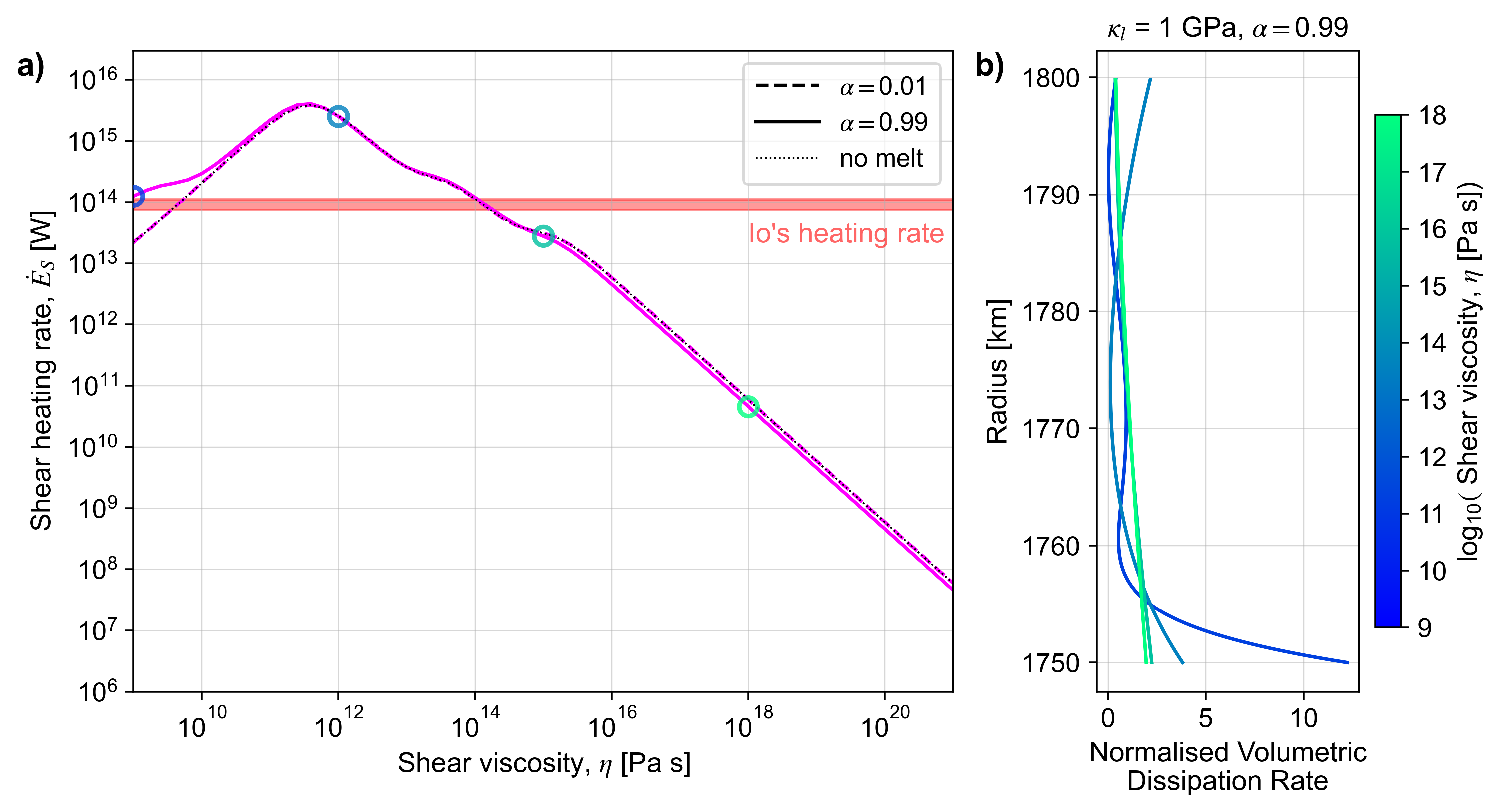}
	\caption{a) Global- and time-averaged shear dissipation rate in the solid, $\dot{E}_{S}$, as a function of shear viscosity, $\eta$, for two different Biot's coefficient $\alpha$. The black dotted line indicates the heating rate for a model with an entirely solid asthenosphere. b) Normalised volumetric heating rate depth profiles for $\alpha=0.99$ and $\kappa_l =$~\SI{1}{\giga\pascal}, which correspond to the open circles in panel a). Normalisation is taken relative to the mean heating rate. In both panels, the solid is taken to be elastic in shear ($\eta \rightarrow \infty$), $\zeta$ is taken to be independent of porosity (which is held at $\phi = 0.1$), the asthenosphere thickness is $H =$~\SI{300}{\km}, and mobility is $M_\phi=$~\SI{5e-7}{\square\metre\per\pascal\per\second}.}
	\label{fig:heating_solid_shear_viscosity}
\end{figure}

\subsubsection{Shear heating} \label{sssec:heating_shear}

Here we consider the classic case of heating in the solid skeleton due to shear deformation, but for a partially molten medium. The results are shown in Figure \ref{fig:heating_solid_shear_viscosity}. We assume finite shear viscosity in the solid, $\eta$, infinite compaction viscosity, $\zeta\rightarrow \infty$, and fix the mobility to $M_\phi =$~\SI{5e-7}{\square\metre \per\pascal\per\second}. The liquid bulk modulus has minimal effect on shear heating, so we arbitrarily take $\kappa_l = \SI{1}{\giga\pascal}$. We also show solutions for a one-phase (solid-only) model ($\porosity = \alpha = 0$).

The heating curves in Figure \ref{fig:heating_solid_shear_viscosity}a are minimally altered from the incompressible-phases case shown in Figure \ref{fig:overview}a. Io-like heating rates are obtained on either side of the critical viscosity, for $\eta \sim \SI{e14}{\pascal\second}$ and $\SI{e10}{\pascal\second}$. When the shear viscosity is ultra-low and sub-critical ($\eta \sim \SI{e9}{\pascal\second}$), melt enhances the shear-heating rate significantly when the asthenosphere is weaker to compaction ($\alpha = 0.99$). This enhanced heating rate is accompanied by a sharp increase in heating at the base of the asthenosphere, shown in Figure \ref{fig:heating_solid_shear_viscosity}b. However, for (more realistic) viscosities that exceed the critical value, melt plays a minor role in altering the shear heating rates.

\subsubsection{Compaction heating} \label{sssec:heating_bulk}

In the previous section, we suppressed compaction dissipation by taking $\zeta \rightarrow \infty$. Here, we relax this limit, and explore how compressibility of the melt and compactibility of the asthenosphere alters compaction heating. We force the solid to behave elastically under shear by taking $\eta \rightarrow \infty$, and fix the mobility to $M_\porosity = \num{e-7}$~\si{\metre\squared\per\pascal\per\second}.  

\begin{figure}[!t]
	\centering
	\includegraphics[width=0.85\linewidth]{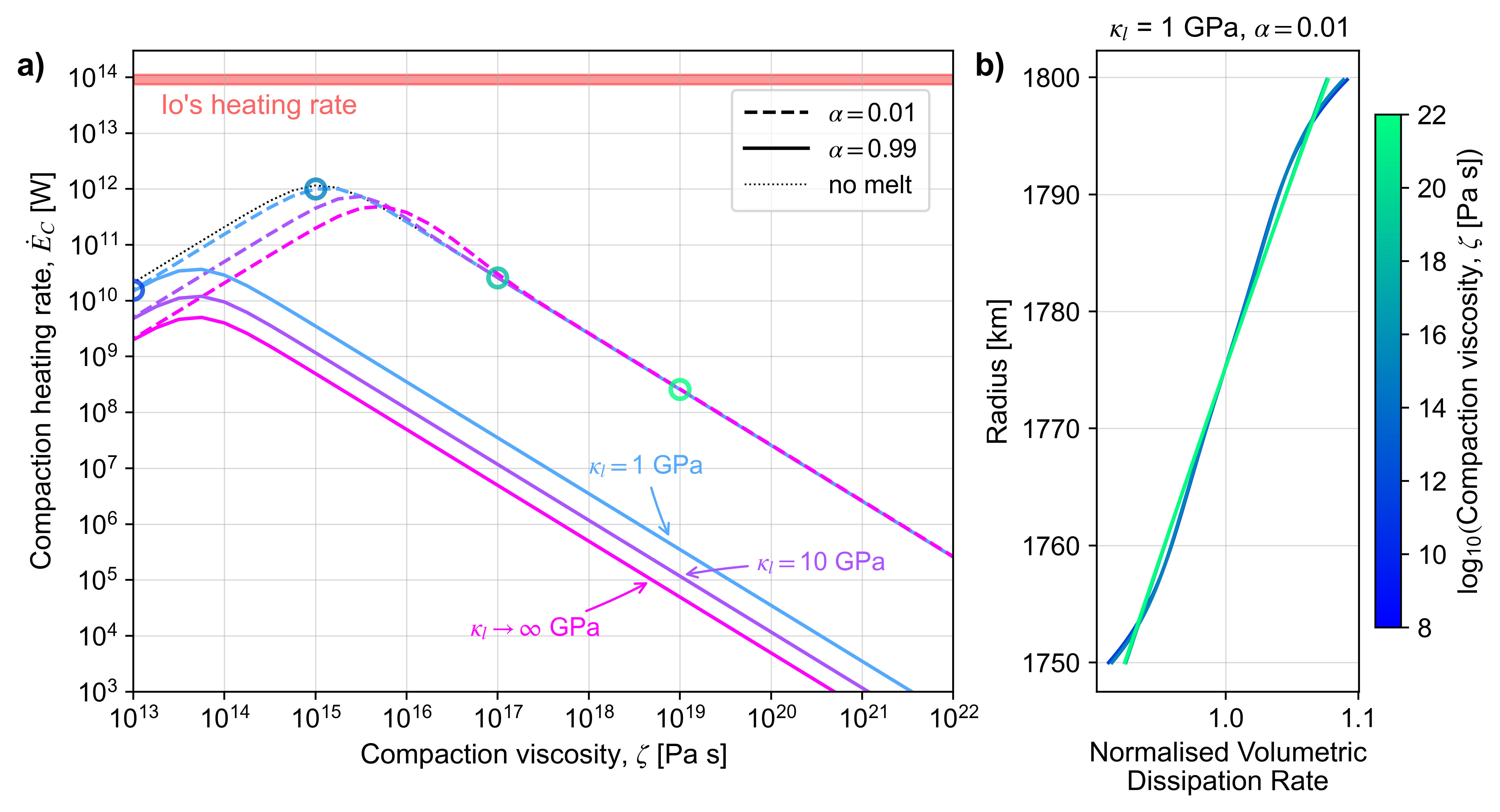}
	\caption{a) Global- and time-averaged compaction dissipation rate in the solid, $\dot{E}_C$, as a function of compaction viscosity, $\zeta$, for different liquid bulk moduli, $\kappa_l$, and Biot's coefficient $\alpha$. b) Normalised- and spherically averaged volumetric heating rate depth profiles for $\alpha=0.01$ and $\kappa_l =$~\SI{1}{\giga\pascal}, which correspond to the open circles in panel a). Normalisation is taken relative to the mean heating rate. In both panels, the solid skeleton is taken to be elastic in shear ($\eta \rightarrow \infty$), $\zeta$ is taken to be independent of melt fraction (which is held at $\phi = 0.1$), the asthenosphere thickness is $H =$~\SI{50}{\km}, and mobility is $M_\phi =$~\SI{e-7}{\square\metre\per\pascal\per\second}.}
	\label{fig:heating_solid_bulk_viscosity}
\end{figure}

The total compaction dissipation rate is shown in Figure \ref{fig:heating_solid_bulk_viscosity}a. The critical compaction viscosity at which compaction dissipation is maximised is sensitive to $\alpha$ but insensitive to $\kappa_l$. Smaller $\alpha$ (stronger matrix) pushes the critical compaction viscosity to larger $\zeta$. The maximum possible bulk heating rate for this interior structure is roughly \SI{1}{T\watt} ($\sim$1\% of Io's observed heating), occurring when $\zeta \sim$~\num{e15}--\SI{e16}{\pascal\second} and $\alpha = 0.01$, which is broadly in agreement with ref.~\cite{kervazo2021SolidTides} as discussed further in Section \ref{sec:discussion}. Significantly smaller heating rates are obtained when the solid matrix is weak (compressible compared to the solid grains ($\alpha = 0.99$)). Although not shown, compaction heating rates that are 10\% of Io's can be reached, but only when $M_\porosity \gtrsim$\num{e-3}~\si{\metre\squared\per\pascal\per\second}.  
Below this value, compaction heating is insensitive to mobility.

Biot's coefficient controls the critical compaction viscosity because it alters the compaction Maxwell time of the asthenosphere.
Maximum compaction dissipation occurs when the tidal forcing period is similar to the compaction Maxwell time, i.e., when $\omega\tau_{C} \sim$$0.1$--$1$. The critical compaction viscosity is therefore $\zeta_{crit} \sim \kappa_d / \omega = (1 - \alpha)\kappa_s / \omega$. Hence, when $\alpha \ll 1$, the critical compaction viscosity is dictated by the compressibility of only the solid grains. As the solid grains become more incompressible, $\alpha \rightarrow 1$, so the critical compaction viscosity required to reach maximum heating decreases, which is the behaviour shown in Figure \ref{fig:heating_solid_bulk_viscosity}a. 

Figure \ref{fig:heating_solid_bulk_viscosity} shows that the total compaction dissipation rate is dominantly controlled by compressibility of the solid skeleton relative to the solid grains. When the skeleton is strong against compaction (small $\alpha$), higher heating rates can be achieved than when the skeleton is comparatively weak (large $\alpha$). The compaction heating rate is partly controlled by $\Im{\f{\kappa}_d}$ \eqref{eq:diss_bulk}, as this dictates the viscous delay in response to compaction stresses. $\Im{\f{\kappa}_d}$ increases as the solid skeleton becomes increasingly strong ($\alpha \to 0$),
\begin{equation}
	\Im{\f{\kappa}_d} = \frac{\omega \zeta}{1 + \frac{\omega^2 \zeta^2}{(1-\alpha)^2 \kappa_s^2}}.
\end{equation}
The maximum value of $\Im{\kappa_d}$ occurs when $\alpha = 0$. This means that the viscous response to compaction is highest when the solid skeleton's elastic resistance to compaction is at its maximum, relative to the solid grains ($\kappa_d = \kappa_s$).  
Thus, somewhat counterintuitively, compaction dissipation is enhanced when the solid skeleton and solid grains have comparable compressibilities, rather than simply when the skeleton is easily compactible. 

Figure \ref{fig:heating_solid_bulk_viscosity}a also shows that the total compaction heating rate generally increases as the melt becomes more compressible, provided that either $\alpha$ is large or the solid skeleton behaves viscously ($\zeta < \zeta_{crit}$). This trend is controlled by how much the solid skeleton can (de)compact. If the melt is highly compressible, the solid skeleton can compact more before pore pressures become high enough to prevent further deformation. Compaction heating thus generally increases with more compressible melt. 
However, if the skeleton is both strong and behaves elastically under isotropic stresses ($\zeta > \zeta_{crit}$), then the amount of compaction is no longer controlled by the pore pressure, and is instead controlled by the elastic resistance of the skeleton. In this case the melt's compressibility no longer controls the heating rate, which is why all dashed lines in Fig.~\ref{fig:heating_solid_bulk_viscosity}a converge when the asthenosphere behaves elastically (large compaction viscosity).

In summary, compaction dissipation is maximised when (\textit{i}) the skeleton is effective at resisting elastic compaction, $\alpha\ll$~1, and (\textit{ii}) when $\zeta$ is close to its critical value so that $\omega \tau_C\sim$~1.

\subsubsection{Darcy heating}\label{sssec:heating_darcy}

\begin{figure}[!t]
	\centering
	\includegraphics[width=0.85\linewidth]{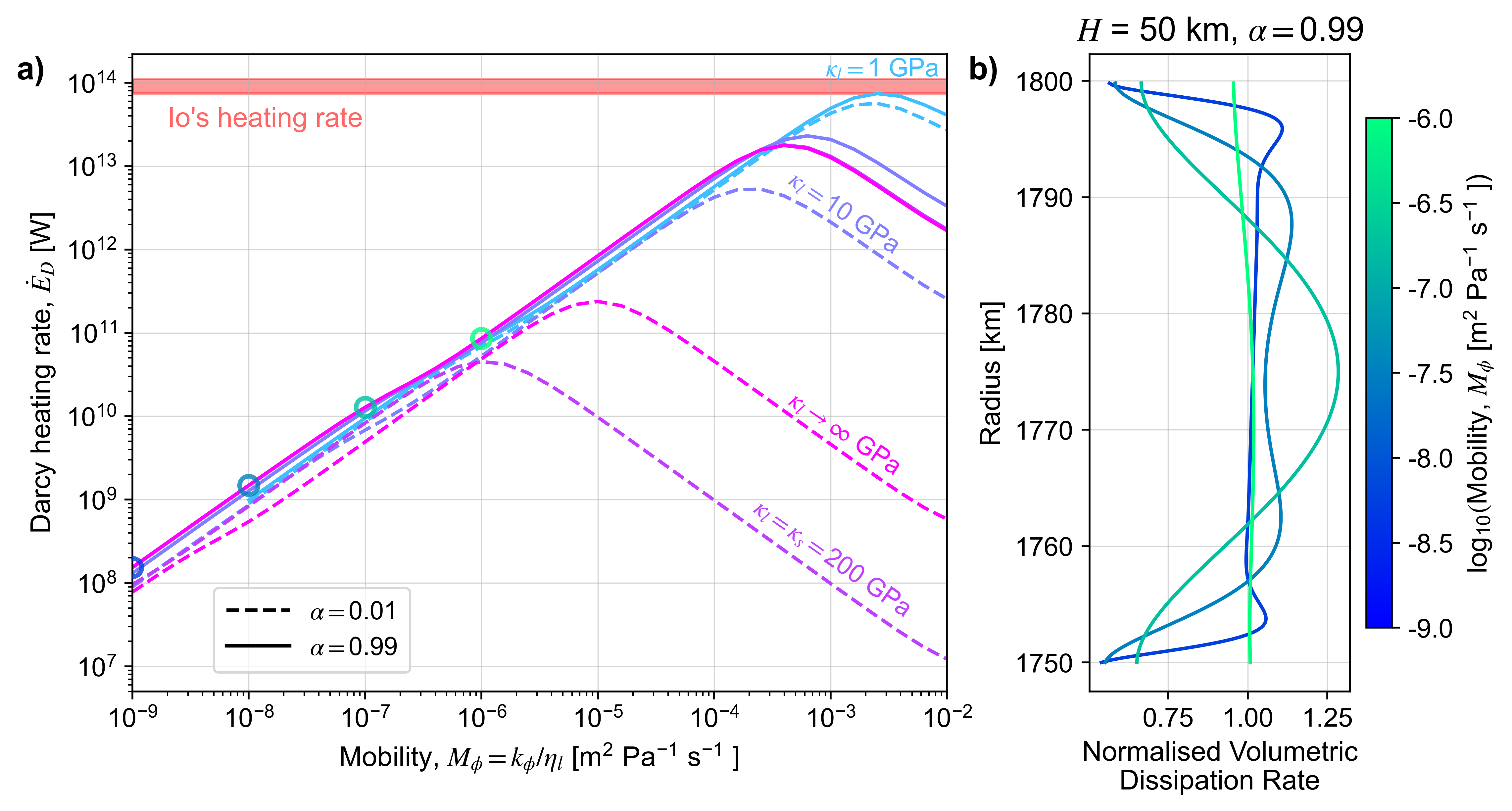}
	\caption{\textbf{a)} Global- and time-averaged Darcy dissipation rate, $\dot{E}_D$, as a function of mobility, $M_{\phi}$, for different liquid bulk moduli, $\kappa_l$, and Biot's coefficient $\alpha$. The colours represent the bulk modulus (compressibility) of the melt, with warmer colours denoting a more incompressible melt. 
		Note that the lower $\kappa_l$ curves do not extend to the smallest mobilities due to numerical instability. \textbf{b)} Normalised- and spherically averaged volumetric heating rate depth profiles, which correspond to the open circles in panel a) for $\kappa_l \rightarrow \infty$. Normalisation is taken relative to the mean heating rate. In both panels, the solid is taken to be elastic, $M_{\phi}$ is taken to be independent of melt fraction (which is held at $\phi = 0.1$), the asthenosphere is $H =$~\SI{50}{\km} thick, and compaction dissipation is ignored.}
	\label{fig:heating_mobility}
\end{figure}

The total Darcy dissipation rate $\dot{E}_{D}$ is given by integrating Equation \ref{eq:diss_liq_mobility} over the volume of the asthenosphere. The result is shown in Figure \ref{fig:heating_mobility}a as a function of the melt mobility $M_\phi$ (Eq.~\ref{eq:mobility}) for a moderately incompressible ($\alpha=0.01$, dashed lines) and compressible ($\alpha=0.99$, solid lines) poroelastic asthenosphere ($\eta\to\infty$, $\zeta \to \infty$).

We see that the critical mobility is sensitive to both $\kappa_l$ and $\alpha$. Maximum heating occurs at higher mobilities as the melt becomes more compressible and the skeleton becomes weaker to compaction. At mobilities less than critical, the Darcy dissipation rate is only weakly sensitive to $\alpha$ and melt compressibility, varying by about a factor of 2--3 between all cases shown. In contrast, $\alpha$ and $\kappa_l$ strongly control the heating rate at mobilities greater than critical. 
The highest possible heating rate occurs for $\kappa_l = \SI{1}{\giga\pascal}$ when the mobility is $\sim \SI{3e-3}{\metre\squared\per\pascal\per\second}$, which can match Io's heat output. For a melt with $\eta_l = $~\SI{1}{\pascal\second}, this gives a permeability of $k =$~ \SI{3e-3}{\square\metre}. Such a high mobility (permeability) can only be attained if the effective grain size is larger than \SI{10}{\cm} (Figure~\ref{fig:param_vs_meltfraction}). This is discussed further in Section \ref{ssec:diss_heating_darcy}.

As recognised in \S\ref{ssec:overview} for incompressible phases, the critical mobility occurs when elastic compaction becomes limited by elastic resistance of the skeleton, rather than the melt's viscous resistance to segregation. When this happens, the segregation flux becomes constant because Darcy flow is also limited by elastic resistance of the skeleton, rather than viscous resistance of the melt. 
When $\alpha$ is small, the skeleton has a greater resistance to elastic compaction, so melt segregation is limited at lower mobilities (i.e., at higher viscous resistance to Darcy flow). The critical mobility, as well as the maximum Darcy heating rate, therefore decreases with $\alpha$. 
Liquid compressibility also affects the critical mobility. If the melt is compressible, then the segregation flux can increase with mobility \textit{past} the point at which compaction of the skeleton is limited by elasticity. The segregation flux is eventually limited when the melt can compress no further. Thus, for compressible melt, there is a decoupling between the mobility at which elastic compaction is halted, and the mobility at which melt segregation is limited. The critical mobility and maximum Darcy heating rate therefore generally increase as the melt becomes more compressible. 
This behaviour is the primary control on the Darcy heating rates shown in Figure \ref{fig:heating_mobility}a.

In Figure \ref{fig:heating_mobility}b, spherically averaged depth profiles of the volumetric Darcy-dissipation rate are shown for a selection of mobilities and $\alpha=0.99$. When the mobility is high (green line), Darcy dissipation is relatively uniform with radius across the asthenosphere. With decreasing mobility (green $\to$ blue lines), boundary layers emerge at the top and bottom of the asthenosphere. These boundary layers lead to numerical instability at small mobilities, as noted by \RNporo and \Kporo. The boundary-layer thickness scales with the compaction length of the asthenosphere, an emergent property of partially molten media \cite{mckenzie1984GenerationCompaction, katz2022DynamicsPartially}.

\subsection{Total heating rate as a function of melt fraction}\label{ssec:total_heating_vs_melt_fraction}

In the previous sections, rheological parameters were imposed independent of the melt fraction $\porosity$. In reality, the permeability of the interconnected pores, compaction/shear viscosity and drained bulk modulus all depend on $\porosity$. In this section we account for these dependencies, allowing us to gain a more physically consistent insight into how melt fraction controls Io's heating rate.

\begin{figure}[t]
	\centering
	\includegraphics[width=\linewidth]{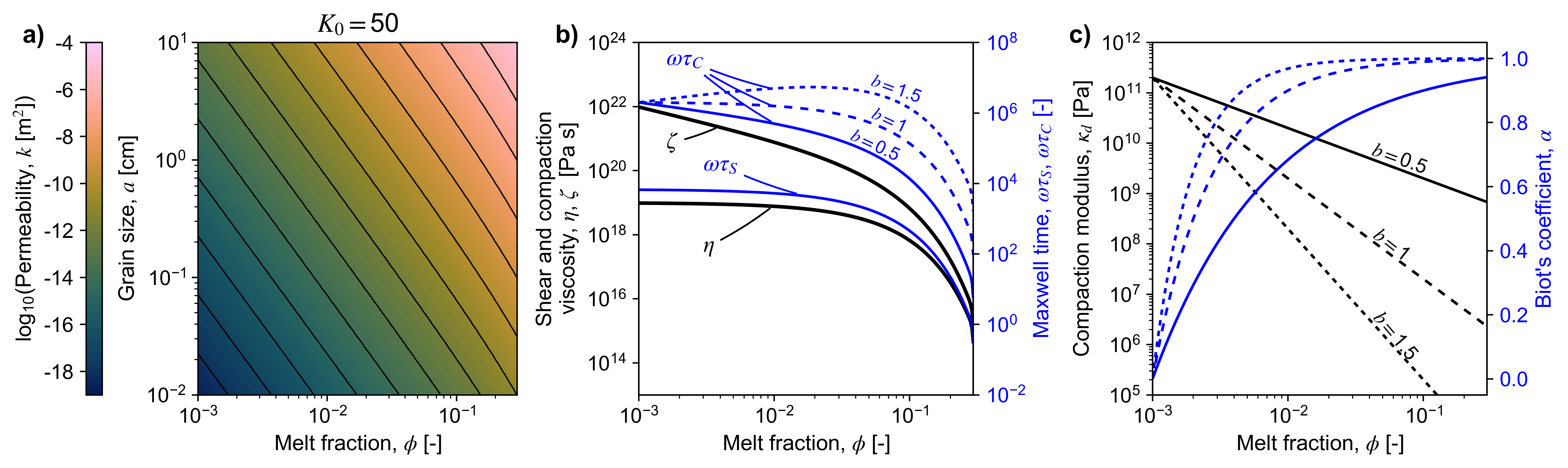}
	\caption{\textbf{a)} Asthenosphere permeability as a function of melt fraction $\porosity$ and grain size $a$, calculated using Eq.~\ref{eq:permbeability_phi} with $K_0 = 50$. For the liquid viscosity of $\eta_l =$~\SI{1}{\pascal\second} used in the manuscript, the colour scale can equivalently be interpreted in terms of mobility $\log_{10}(M_\phi)$. \textbf{b)} Compaction and shear viscosity ($\zeta$ and $\eta$, black), and the corresponding nondimensional Maxwell times ($\omega \eta/ \mu$ and $\omega \zeta / \kappa_d$, blue), as a function of melt fraction, calculated with Eqs. \ref{eq:bulk_viscosity_phi} and \ref{eq:shear_viscosity_phi}. \textbf{c)} Drained bulk modulus (black) and Biot's coefficient (blue) as a function of melt fraction, calculated with Eqs.~\ref{eq:bulk_modulus_phi} and \ref{eq:biot_coeff_phi}. The power-law exponent used to calculate $\kappa_d$ in panels \textbf{b)} and \textbf{c)} is set to $b=$~0.5 (solid lines), 1 (long dashes), and 1.5 (short dashes). 
	}
	\label{fig:param_vs_meltfraction}
\end{figure}

We assume that permeability scales with porosity as \cite{dullien2012PorousMedia, mckenzie1984GenerationCompaction, Rudge2018co},
\begin{equation}\label{eq:permbeability_phi}
	k(\phi) =  \frac{a^2 \phi^3}{K_0 (1-\porosity)^2},
\end{equation}
\noindent where $a$ is the radius of the solid grains and $K_0\approx 50$ is an empirically determined constant \cite{bear2013dynamics}. Figure \ref{fig:param_vs_meltfraction}a shows the permeability for different melt fractions and grain sizes using $K_0 = 50$. The highest permeability of $k = \SI{e-5}{\square\metre}$ occurs for the largest grain size of \SI{10}{\cm} at melt fractions of $\phi\sim 0.3$. For Earth-like mantle melt fractions of $\phi\sim 0.01$ and grain sizes of \SI{1}{\mm}, permeability is substantially smaller at $k \sim$~\SI{e-12}{\square\metre}.  

The compaction viscosity of Earth's mantle has never been directly measured \cite{katz2022physics}. Only ref.~\cite{renner2003MeltExtraction} has experimentally inferred $\zeta$ from compaction rate experiments on peridotite samples. Yet its role is essential in a compacting poro-viscoelastic material. Assuming that compaction is associated with viscous deformation of grains during the closure/expansion of pores in partially molten rock, it can be shown that the compaction viscosity should vary as \cite{batchelor1967IntroductionFluid, hewitt2008PartialMelting}
\begin{equation}
	\label{eq:bulk_viscosity_phi}
	\zeta(\porosity) \approx \frac{\eta}{\phi},
\end{equation}
provided that the porosity is small. The compaction viscosity is therefore a property of the solid--liquid aggregate, rather than the solid grains. Also following refs.~\cite{kervazo2021SolidTides, costa2009ModelRheology}, we modify the shear viscosity as a function of melt fraction through
\begin{equation}\label{eq:shear_viscosity_phi}
	\eta(\porosity) = \eta_l\frac{1 + \Theta^\ell }{\left[1 - F(\Theta, \xi, \gamma)\right]^{5(1-\porosity_*)/2}},
\end{equation}
where the additional auxiliary functions are
\begin{align}
	\Theta &= \frac{1-\porosity}{1-\porosity_*}, \\
	F(\Theta, \xi, \gamma) &= (1- \xi) \text{erf} \left(\frac{\sqrt{\pi}}{2(1-\xi)} \Theta (1+\Theta^\gamma)\right).
\end{align}
The other rheological parameters are determined experimentally in ref.~\cite{costa2009ModelRheology}, with $\ell = 25.7$, $\porosity_* = $~0.431, $\xi = \num{1.17e-9}$, and $\gamma = 5$, which we take from Table 3 in ref.~\cite{kervazo2021SolidTides} (noting the typo in their $\phi^*$, see ref.~\cite{kervazo2022InferringIos}).  The function in Eq.~\ref{eq:shear_viscosity_phi} accounts for the decrease of the shear viscosity due to grain-boundary lubrication by the melt. Hence, $\eta$ is a property of the two-phase aggregate, and not a property of the solid grains.

Unlike ref.~\cite{kervazo2021SolidTides}, we keep $\mu$ constant. There is no experimental determination of how the elastic shear modulus should vary with melt fraction, but it is expected to be generally insensitive to $\porosity$ except near the disaggregation limit \cite{mckenzie1984GenerationCompaction}. Near disaggregation, it is questionable whether the rheological and two-phase model used here is valid. The bulk modulus of the solid grains $\kappa_s$ is a property of the grains, not the two-phase aggregate, and so is independent of melt fraction.

Figure \ref{fig:param_vs_meltfraction}b plots Eqs.~\ref{eq:bulk_viscosity_phi} \& \ref{eq:shear_viscosity_phi} for $\zeta$ and $\eta$ as a function of melt fraction. The shear viscosity stays roughly constant until $\porosity \sim 0.05$, above which it decreases to about $\eta \sim \SI{e15}{\pascal\second}$ at the disaggregation limit. The compaction viscosity decreases steadily as melt fraction increases, again until $\porosity \sim 0.05$, at which point it rapidly decreases to $\zeta \approx 3\eta$ at the disaggregation limit. We also plot the Maxwell times for shear and compaction as a function of $\porosity$. For shear, $\omega \tau_S \sim 1$ very close to the disaggregation melt fraction. For compaction, $\omega \tau_C \sim 1$ again when close to disaggregation, but only for $b \lesssim 0.5$. Significant compaction dissipation is therefore favoured for small $b$.

Finally, the compaction modulus of the solid skeleton is also controlled by melt fraction \cite{selvadurai2020InfluencePore}. The relationship between drained compaction modulus and melt fraction can be expressed in terms of another unknown rheological constant, the porosity bulk modulus (see Eq.~4.159 in ref.~\cite{cheng2016Poroelasticity}), which determines the skeleton's resistance to grain rearrangement. For simplicity, we instead adopt a power law relationship between $\kappa_d$ and $\porosity$. As the melt fraction approaches zero, we expect that $\kappa_d \to \kappa_s$, because the solid skeleton adopts the properties of the grains themselves. We also expect $\kappa_d$ to go to zero as $\porosity \rightarrow \porosity_\text{crit}$, as there is then no coherent solid skeleton to resist isotropic stresses. This behaviour can be approximated by
\begin{equation}\label{eq:bulk_modulus_phi}
	\kappa_d(\porosity) = 
	\begin{cases}
		\kappa_s & \text{if $\,\porosity < \porosity_0$} \\
		\kappa_s \left( \frac{\porosity_0}{\porosity}\right)^b & \text{if $\,\porosity_0 \leq \phi < \porosity_{\text{crit}}$} \\
		0 & \text{if $\,\porosity \geq \porosity_{\text{crit}}$}
	\end{cases} 
\end{equation}
where $b>0$ is an unknown constant, and $\phi_0$ is a small (but non-zero) melt fraction below which the skeleton behaves like a coherent solid, $\kappa_d = \kappa_s$. Correspondingly, Biot's  coefficient becomes,
\begin{equation}\label{eq:biot_coeff_phi}
	\alpha (\porosity) = 
	\begin{cases}
		0 & \text{if $\, \porosity < \porosity_0$} \\
		1 - \left( \frac{\porosity_0}{\porosity}\right)^b & \text{if $\, \porosity_0 \leq \phi < \porosity_{\text{crit}}$ } \\
		1 & \text{if $\, \porosity \geq \porosity_{\text{crit}}$ }
	\end{cases}.
\end{equation}
We adopt $\porosity_0 = \num{e-3}$, and explore a range of $b$ values. Both $\kappa_d$ and $\alpha$ are shown as a function of melt fraction in Figure \ref{fig:param_vs_meltfraction}c for $b=\{0.5,\, 1.0,\, 1.5\}$. Increasing melt fraction causes $\alpha$ to approach unity and $\kappa_d$ to decrease. The increase in $\alpha$ and drop in $\kappa_d$ is more rapid as $b$ becomes larger. For $b \ll 1$, we always have $\alpha \ll 1$, even close to $\phi_{\text{crit}}$. Thus, small $b$ gives the asthenosphere a strong elastic resistance against compaction across the full range of relevant melt fractions, $0 < \phi < \phi_{\text{crit}}$.

The shear viscosity formulation in Eq.~\ref{eq:shear_viscosity_phi} is chosen to enable direct comparison with results from ref.~\cite{kervazo2021SolidTides}. This formulation was developed to capture the increase in effective shear viscosity as interactions between suspended crystals begin to dominate the magma's rheology \cite{costa2009ModelRheology}. Alternative formulations are derived in terrestrial mantle deformation studies, where the microphysical mechanism that enables irreversible deformation gives rise to its own shear viscosity law. For instance, at sufficiently large grain sizes and/or high stress, dislocation creep is the dominant deformation mechanism \cite{kohlstedt1974LowstressHightemperature}. However, at Io's high melt fraction, diffusion-accommodated grain-boundary sliding may instead be dominant \cite{hirth1995ExperimentalConstraintsa}. Future work could explore the consequences of these different deformation mechanisms.

\subsubsection{Heating rates}

\begin{figure}[!t]
	\centering
	\includegraphics[width=0.95\linewidth]{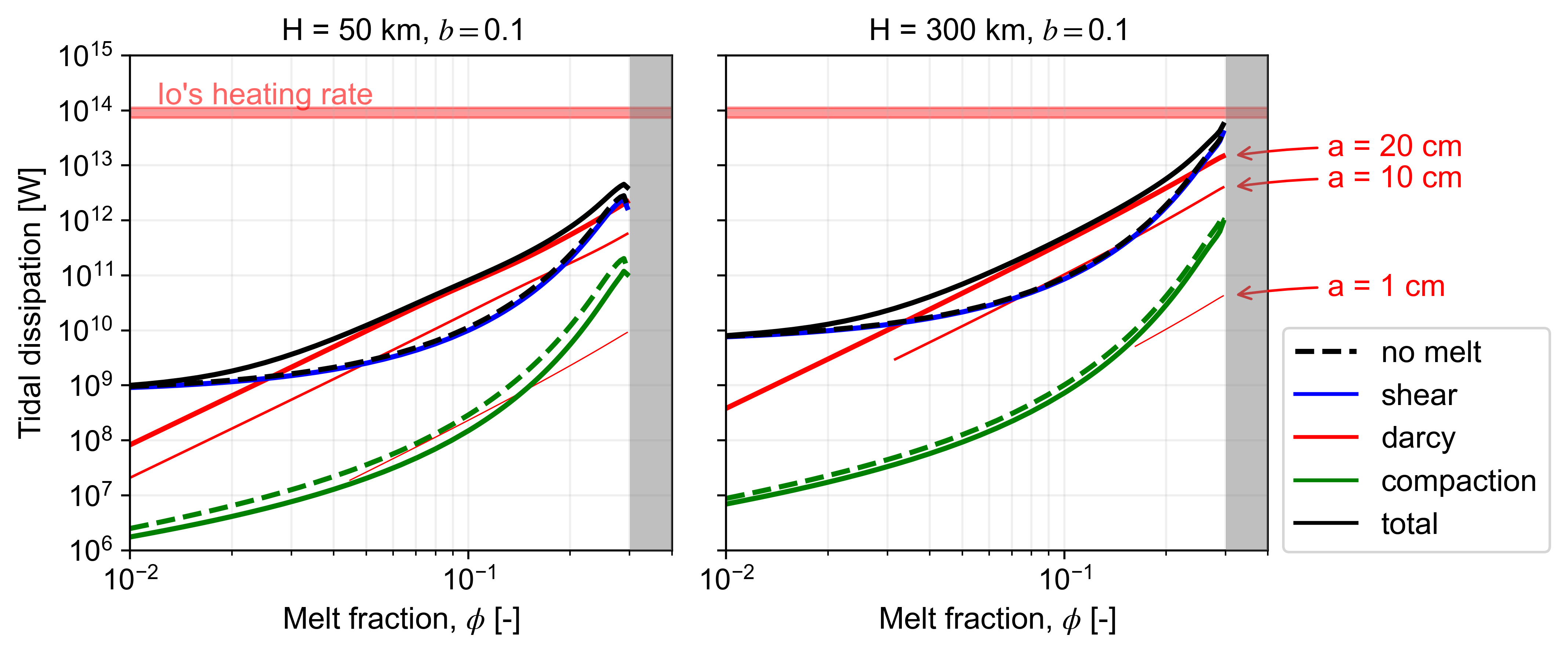}
	\caption{Tidal dissipation rate as a function of melt fraction for an asthenosphere thickness of $H=\SI{50}{\km}$ (left) and \SI{300}{\km} (right). Solid lines are solutions computed for the two-phase problem, while dashed lines are solutions that neglect two-phase dynamics. The different colours represent dissipation from shear (blue), compaction (green), and Darcy (red) deformation. Compaction dissipation without melt (green dashed lines) is calculated for a single phase model where $\alpha=0$ and $\phi=0$, though $\zeta$ remains a function of $\porosity$, as in ref. \cite{kervazo2021SolidTides}. As labelled on the right, the different thickness of red line correspond to grain sizes of  $a = \SI{1}{\cm}$ (thin), \SI{10}{\cm} (medium), and \SI{20}{\cm} (thick). The black line is the total dissipation for no melt (dashed) and largest grain size (solid). Curves do not extend across all melt fractions due to numerical instabilities encountered at small melt fractions/mobilities.}
	\label{fig:heating_melt_fraction}
\end{figure}

In the following, we modify $k$, $\zeta$, $\eta$, $\alpha$, and therefore $\kappa_d$, as functions of melt fraction using Eqs.~\ref{eq:permbeability_phi}, \ref{eq:bulk_viscosity_phi}, \ref{eq:shear_viscosity_phi}, \ref{eq:biot_coeff_phi}, and \ref{eq:bulk_modulus_phi}, respectively. We take $\mu$, $\kappa_s$, and $\kappa_l$ to all be independent of melt fraction.

Figure \ref{fig:heating_melt_fraction} shows the different contributions of Darcy, compaction, and shear dissipation to the total heating rate of our Io model. Darcy dissipation is substantially greater than shear dissipation across the middle range of melt fractions ($\porosity \sim 0.05$--$0.15$), provided that the grain size is $a\gtrsim\SI{10}{\cm}$. For grain sizes that are less than $\SI{1}{\cm}$, it is at least two orders of magnitude less than the shear-heating rate at any melt fraction. The critical mobility is never encountered for any melt fraction. In all cases, the total dissipated power increases as the asthenosphere thickness is increased from $H=\SI{50}{\km}$ (left) to \SI{300}{\km} (right), largely due to the increase in heat-producing volume.

Two-phase dynamics generally decreases both shear and compaction heating rates, as compared to the single-phase results. For shear heating, this decrease is minor. For compaction dissipation, Figure \ref{fig:heating_melt_fraction} shows that this decrease is also minor, but this result is highly sensitive to the exponent $b$.
As discussed around Figure \ref{fig:heating_solid_bulk_viscosity}, there are two requirements to maximising compaction dissipation: (\textit{i}) the non-dimensional compaction Maxwell time of the asthenosphere must satisfy $\omega \tau_C \sim$~0.1--1 (i.e., $\zeta$ must be close to its critical value), and (\textit{ii}) the solid matrix must be (relatively) elastically strong against compaction stresses, meaning $\alpha$ is small. The latter requirement can easily be satisfied in the one-phase model because $\alpha = 0$. To what degree these two requirements are met in the two-phase model depends on how sensitively $\alpha$ depends on melt fraction. Here, this sensitivity depends on the unknown exponent $b$ through Eq.~\ref{eq:biot_coeff_phi}.

\begin{figure}[t]
	\centering
	\includegraphics[width=1.0\linewidth]{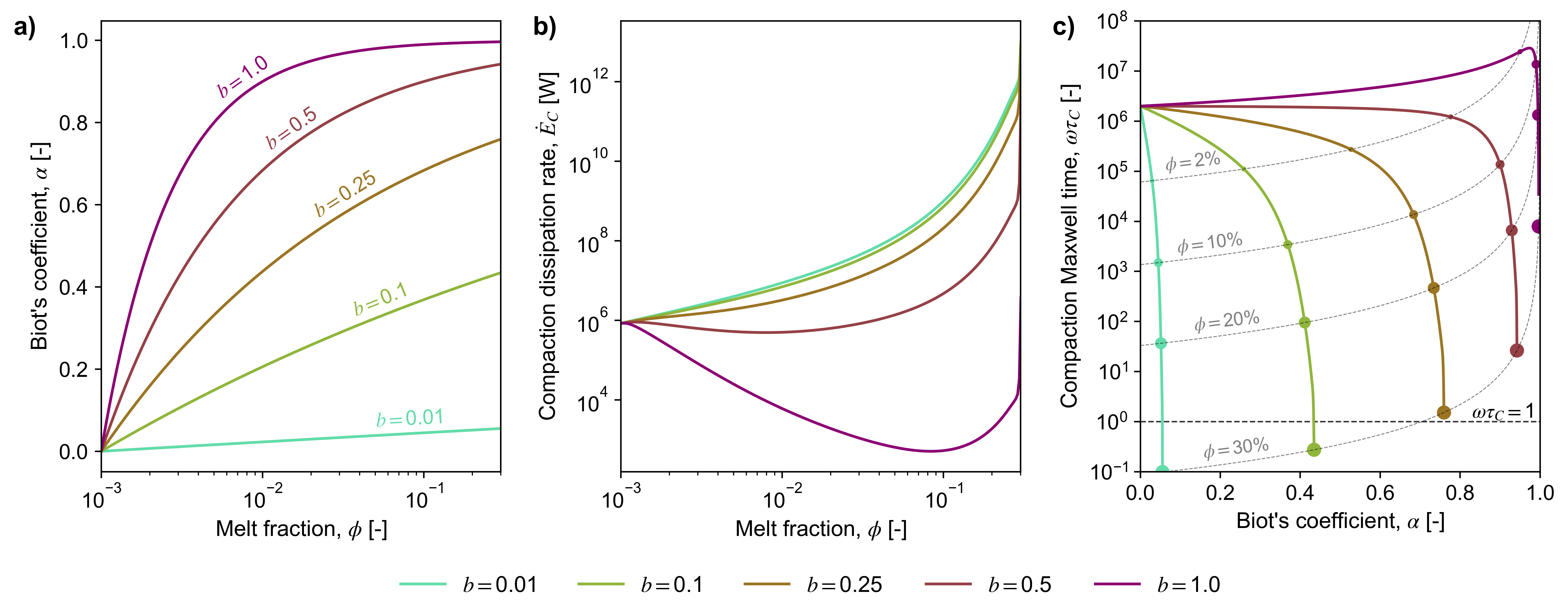}
	\caption{Impact of different $b$ exponents on \textbf{a)} Biot's coefficient and \textbf{b)} compaction dissipation rate as  a function of melt fraction, and \textbf{c)} compaction Maxwell time $\omega \tau_C$ as a function of Biot's coefficient $\alpha$. Also plotted in c) are contours of melt fraction, and circles indicate where these intersect the Maxwell time curves for a given $b$.  Panel c) should thus be interpreted as representing the trajectory of $\alpha$ and $\omega \tau_C$ across the full range of melt fractions, for a given $b$.  In this figure we have forced $k=$~\SI{e-6}{\square\metre} to aid numerical stability. The compaction dissipation rate is insensitive to $k$ unless the mobility of the asthenosphere is $\gtrsim$~\SI{e-3}{\square\metre\per\pascal\per\second}.}
	\label{fig:params_vs_alpha}
\end{figure}

The relationship between Biot's coefficient and compaction dissipation as a function of $b$ and $\porosity$ is shown in Figure \ref{fig:params_vs_alpha}. In panel a) we see that the range of permitted $\alpha$ values (constrained by $\porosity_0 < \porosity < 0.3$) becomes increasingly small as $b$ decreases. Hence, smaller $b$ promotes an asthenosphere that can more effectively resist elastic compaction. Panel b) shows that the maximum possible compaction dissipation rate is only reached as $b\rightarrow 0$, for any melt fraction. When $b=1$, compaction dissipation peaks at small and large melt fractions, but is much weaker at intermediate melt fractions. When $b \lesssim 0.5$, compaction dissipation generally increases with melt fraction. 
Panel c) shows how the compaction Maxwell time $\omega \tau_C$ varies with $b$, Biot's coefficient, and melt fraction. We see that, as $b$ decreases, $\omega \tau_C$ approaches unity at smaller $\alpha$ (stronger asthenosphere). Only for the largest melt fraction of $\phi=0.3$ and smallest $b=0.01$, does the asthenosphere reach $\omega \tau_C =$~0.1 and simultaneously have $\alpha \ll 1$. 
Thus, the two conditions to maximise compaction dissipation are more easily met for small $b$ and high melt fraction. Physically, this means that for our self-consistent treatment of melt, the significance of compaction dissipation becomes a question of how quickly the solid skeleton becomes compactible (relative to the solid grains) after the onset of melting. When $b=1$, $\alpha$ is very sensitive to melt fraction so the skeleton becomes rapidly compactible with increasing $\porosity$. In contrast, for $b=0.01$, $\alpha$ is insensitive to $\phi$, so the matrix remains highly uncompactible across the full range of possible melt fractions. This explains the main difference between our two-phase results and the single-phase results of ref.~\cite{kervazo2021SolidTides}, where they implicitly assume $\alpha = 0$. We discuss this point further in \S\ref{sec:discussion}.

\subsection{Distribution of heating} \label{ssec:patterns}

\begin{figure}[!t]
	\centering
	\includegraphics[width=0.6\linewidth]{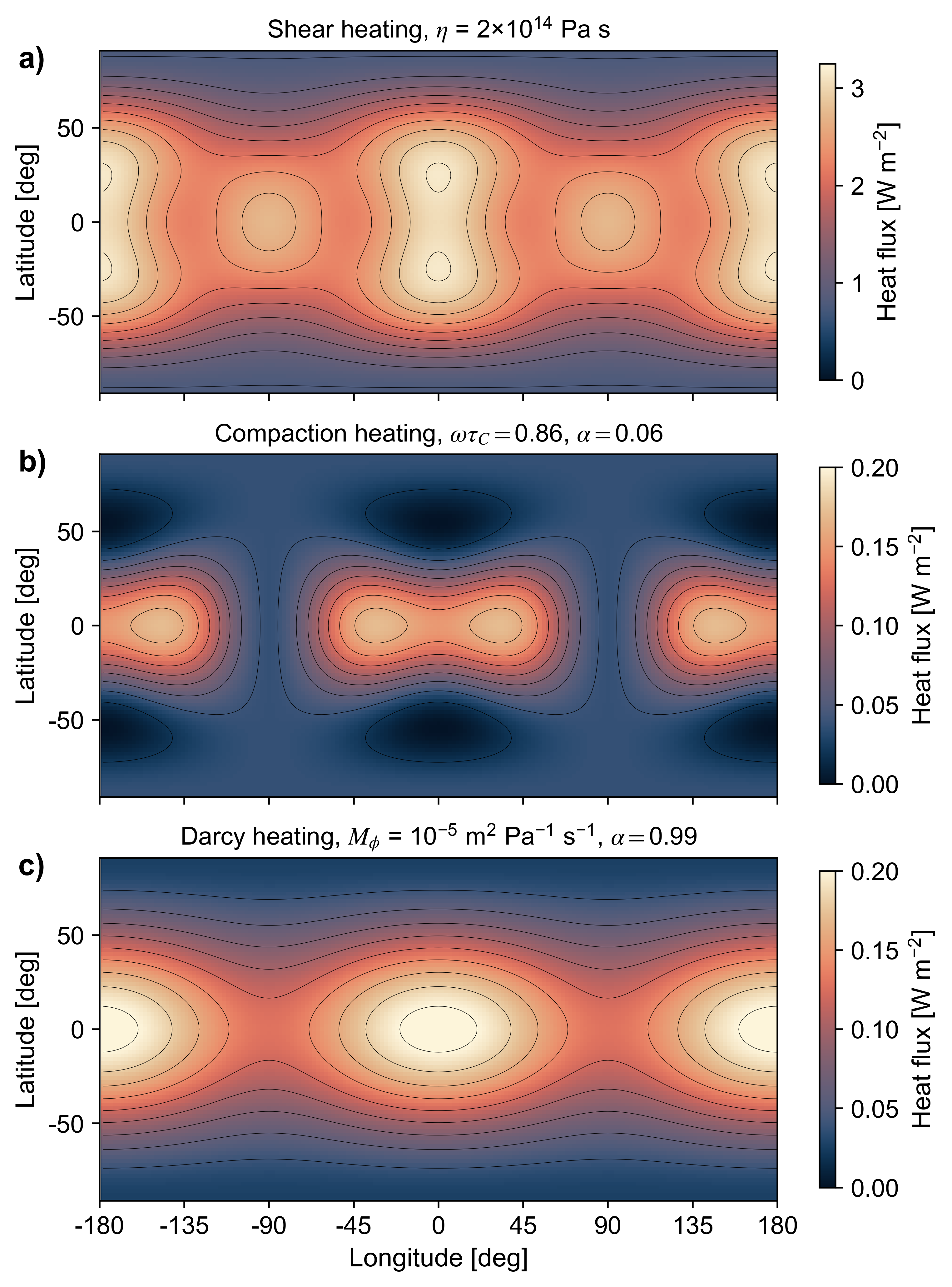}
	\caption{Surface heat flux patterns due to \textbf{a)} shear, \textbf{b)} compaction, and \textbf{c)} Darcy deformation. These patterns are calculated with optimum parameters to maximise heating from each deformation mechanism. Darcy dissipation is calculated for a high mobility of \SI{e-5}{\metre\squared\per\pascal\per\second}, corresponding to a grain size of $a \sim \SI{10}{\cm}$ and melt fraction of $\porosity \sim 0.3$using Eq.~\ref{eq:permbeability_phi}. Compaction dissipation is calculated for optimum viscoelastic parameters at high melt fraction such that $\alpha\ll1$ and $\zeta \sim$~\SI{3.6e15}{\pascal\second}, and shear dissipation uses a low shear viscosity of $\eta =$~\SI{2e14}{\pascal\second}.}
	\label{fig:patterns}
\end{figure}

A key observable property of Io is the distribution of volcanic activity at its surface. If the spatial dependence of tidal heat generation in its interior reflects the distribution of volcanic activity---a significant assumption---then volcanic activity provides a possible path to distinguishing between different tidal-heating modes. Hence, there is a history of comparing Io's volcanic distribution to patterns of tidal heat fluxes predicted by various models of tidal deformation and internal structures \cite{bierson2016TestIo, rathbun2018GlobalDistribution, matsuyama2022TidalHeating, aygun2024TidalHeating}. Imaging of Io by the Galileo and Juno spacecraft have now provided near-full coverage of the global distribution of hotspots. However, there is still active debate about the exact interpretation of this data. Ref.~\cite{zambon2023IoHot} finds that there is a greater concentration of hotspots towards the poles. Refs.~\cite{davies2024IosPolar} and \cite{davies2024NewGlobal} find that while there are, on average, the same areal density of volcanic hotspots across Io, those at lower latitudes have a greater thermal power emission. They argue that this is consistent with models considering solid-only heating in a shallow asthenopshere. Ref.~\cite{pettine2024JIRAMObservations} used a spherical harmonic decomposition of the hotspot distribution from 11 Juno flybys, finding that the activity is not strongly correlated to any of the solid-only or magma-ocean tidal heating models (e.g., ref.~\cite{matsuyama2022TidalHeating}). Below, we compare the heating patterns predicted by the three deformation modes explored here, shear, compaction, and Darcy deformation, shown in Figure \ref{fig:patterns} for a $H=\SI{300}{\km}$ thick asthenosphere.

All three heating patterns shown in Figure \ref{fig:patterns} focus heating to low latitudes. Shear-generated heat flux is shown in panel a), which shows a classic ``shallow'' asthenosphere heating pattern \cite{segatz1988TidalDissipation}, where heating is maximised at low latitudes, at the sub- and anti-Jovian longitudes ($\varphi=\SI{0}{\degree}$ and $\ang{180}$, respectively), and about $\pm \SI{25}{\degree}$ above and below the equator. Our inclusion of two-phase flow does little to change this classic result. Compaction dissipation (panel b) has a heat flux of $\sim\SI{0.2}{\watt\per\square\metre}$, with maxima occurring either side of the sub- and anti-Jovian points, similar to that found by ref.~\cite{kervazo2021SolidTides} (their Fig.~7). Regions of effectively zero heat flux occur north and south of these points. Darcy dissipation, shown in c) for a high-mobility asthenosphere of $M_\porosity = \SI{e-5}{\metre\squared\per\pascal\per\second}$, has dissipation maximised along the equator. The heat-flux magnitude is similar to that of compaction, and peaks at the sub- and anti-Jovian points. There is no longitudinal offset in peak heating, with minimum but non-zero heating at the poles. Overall, however, the broadest scales of heating are similar between compaction and Darcy dissipation.

Overall, our calculations suggest that all heating modes, if located in the asthenosphere, focus heating to low latitudes. Shear deformation causes the highest heating rates if Io's mantle's effective grain size cannot exceed \SI{10}{\cm}. Darcy dissipation heat fluxes can approach that due to shear deformation if the melt's mobility can approach $M_\porosity \sim \SI{e-3}{m^{2}.Pa^{-1}.s^{-1}}$. If tidal heating is directly correlated with the locations of surface volcanism on Io, then the inclusion of two-phase dynamics and physically consistent compaction dissipation does not change the general anticipation that shallow mantle tidal heating is focused towards low latitudes.

\section{Discussion} \label{sec:discussion}

We have shown with our two-phase mantle deformation model that melt segregation and viscous compaction of the two-phase assemblage can generate additional sources of heating in Io's interior. The exact proportion of Io's heating that is controlled by these mechanisms, relative to shear deformation, depends on the rheological properties of the asthenosphere. In \S\ref{ssec:diss_heating_darcy} and \S\ref{ssec:diss_heating_bulk} we discuss how these properties limit the upper bounds of Darcy and compaction dissipation. We identify the limitations and future avenues of this work in \S\ref{ssec:diss_limitations_future}.

\subsection{Magnitude of Darcy dissipation}\label{ssec:diss_heating_darcy}

The magnitude of Darcy dissipation is primarily controlled by the asthenosphere's mobility $M_\porosity$, the ratio of permeability $k$ to melt viscosity $\eta_l$ (Eq.~\ref{eq:mobility}). As shown in Figure \ref{fig:heating_mobility}, Io-like heating rates can be achieved for a high mobility of $M_\porosity \sim \SI{1e-3}{\square\meter\per\pascal\per\second}$. If the melt viscosity is $\eta_l = \SI{1}{\pascal\second}$, then this requires a permeability of $k=\SI{e-3}{\square\metre}$. Our results in \S\ref{ssec:total_heating_vs_melt_fraction} set the permeability using the classic Kozeny-Carman relationship (Eq.~\ref{eq:permbeability_phi}), and Figure \ref{fig:param_vs_meltfraction}a shows that $k \sim \SI{1e-3}{\square\metre}$ can only be achieved at grain sizes of $a > \SI{10}{\cm}$. If the Kozeny-Carman relationship is still valid at such large grain sizes, then the question becomes, what is the grain size in Io's mantle? Grain sizes in Earth's deep mantle are poorly known \cite{dannberg2017ImportanceGrain}. In the shallow mantle, observations and modelling suggest millimeter to centimeter-sized grains \cite{hirth2003RheologyUpper, ruh2022GrainsizeevolutionControls}. Fluid alteration can potential yield grains sizes $>$\SI{10}{\cm} \cite{dannberg2017ImportanceGrain}, though Io's interior is thought to be volatile depleted \cite{dekleer2024IsotopicEvidence}. Modelling of Earth's Moon's tidal response suggests grain sizes of around \SI{1}{\cm} \cite{nimmo2012DissipationTidal}. We therefore conclude that grain sizes in Io's mantle are unlikely to exceed $\sim$\SI{10}{\cm}.

Alternatively, we could interpret a high mobility as reflecting a large \textit{effective} permeability (i.e., averaged over some large area or volume). This could arise from, for example, channelisation of magma within Io's asthenosphere. Magmatic channelisation in Earth's asthenosphere is evidenced by tabular dunite zones found in ophiolites \cite{quick1982OriginSignificance}. These channels are suspected to form via a reactive flow instability, where pressure-driven undersaturation in the magma forces a melting reaction, dissolving pyroxene and precipitating olivine from the melt onto the solid residuum \cite{aharonov1995ChannelingInstability, jones2018ReactioninfiltrationInstability}. The permeability of such channels would be orders of magnitude higher than the surrounding mantle, potentially large enough for turbulent dissipation. However, without observation of such structures on Io, this idea is only speculative.

If effective grain sizes of $\sim$\SI{10}{\cm} are possible, Darcy dissipation can exceed or match the shear-heating rate in the solid when the melt fraction is $0.05 < \porosity < 0.2$, depending on the asthenosphere thickness and melt viscosity (Fig.~\ref{fig:heating_melt_fraction}). This conclusion hinges on the assumption that the asthenosphere behaves as a poro-viscoelastic \textit{Maxwell} material, which we have used here. We know that in general, Maxwell viscoelastic models do not accurately predict attenuation rates that are measured from laboratory experiments across a range of frequencies \cite{mccarthy2016TidalDissipation, bierson2024ImpactRheology}. The Andrade rheological model has been proposed as an alternative that better predicts attenuation in solid, cold, high-viscosity rock and ice \cite{castillo-rogez2011TidalHistory}. However, while there are measurements of the attenuation behaviour of partially molten rock across seismic frequencies at small melt fractions ($\phi \sim 0.01$, \cite{jackson2004ShearWave}), there are, to our knowledge, no such measurements close to tidal frequencies and at large ($\phi \sim 0.1$) melt fractions. Moreover, Andrade model parameters have only been measured for shear deformation, while here we also consider isotropic deformation. It may be the case that an Andrade-type rheology is applicable to a partially molten asthenosphere, but it is not clear what the model's parameters should be when a high melt fraction, two-phase flow, and compaction are included. This is not to say that an Andrade rheology is inappropriate to our case here, but merely that caution should be exercised when extending the model to additional deformation modes. We find, though, that two-phase flow dynamics has only a small impact on shear deformation of the solid skeleton. This small impact potentially indicates that, even with the caveats above, an Andrade model could still be an acceptable rheological description of shear deformation in a partially molten solid at high melt fraction.  

It is also possible that Io's magma may be less viscous than the \SI{1}{\Pa\second} assumed here. There is a general lack of evidence for chemically evolved, high-viscosity magmas on Io \cite{keszthelyi2001ImagingVolcanic}, with lava flow run-out distances and eruption temperatures consistent with low-viscosity, ultramafic melt \cite{mcewen1998HighTemperatureSilicate,keszthelyi2007NewEstimates}. Experimental determination of Earth-like upper-mantle basalt melts yield viscosities as low as $\sim \SI{0.1}{\pascal\second}$ \cite{kushiro1986ViscosityPartial}. Laboratory measurements of peridotite melt have determined that viscosities of even \SI{0.01}{\pascal\second} are possible, though this requires slightly greater pressures than in the asthenosphere that we assume here \cite{liebske2005ViscosityPeridotite}. Low viscosity melt enhances the liquid's mobility, generally increasing the Darcy dissipation rate. If the melt approached $\eta_l \sim \SI{0.01}{\pascal\second}$ in Io's hot asthenosphere, then Darcy dissipation could approach shear dissipation at a more moderate grain size of \SI{1}{\cm}. It should be noted though that measurements of sulphur and chlorine isotopic anomalies in Io's tenuous atmosphere indicate a volatile-depleted interior \cite{dekleer2024IsotopicEvidence}. Volatiles generally decrease the viscosity of magma, so these isotopic measurements could indicate that ultra-low magma viscosities are unlikely at the present day, depending on the magma's silica content.

Darcy (and compaction) heating, while smaller in magnitude than the global shear-heating rate, can still potentially be relevant in driving aspects of Io's internal dynamics at heating rates of \num{e11}-- \num{e12}~W. This heating can be focused at different depths depending on the heating mechanism (Figs.~\ref{fig:overview}--\ref{fig:heating_mobility}), and hence the radial location of buoyancy sources may differ between Darcy, compaction, and shear heating, potentially resulting in different mantle dynamics \cite{tackley2001ThreeDimensionalSimulations}. 

\subsection{Magnitude of compaction dissipation} \label{ssec:diss_heating_bulk}

Ref.~\cite{kervazo2021SolidTides} was the first to present tidal heating estimates for Io that included compaction dissipation. In their work, the asthenosphere was technically treated as a single-phase solid system. The viscoelastic parameters ($\eta$, $\zeta$, $\kappa_s$, $\mu$) all depended on melt fraction, but melt segregation (i.e., two-phase flow dynamics) was neglected. In particular, their $\kappa_s \to \kappa_l$ as $\porosity \rightarrow \porosity_{crit}$. In our two-phase approach, we have necessarily introduced the compaction modulus, $\kappa_d$, which measures the (inverse) compressibility of the solid skeleton when the melt has been drained, rather than the grains themselves. This distinction is important; a two-phase aggregate with incompressible solid and liquid phases ($\kappa_s$, $\kappa_l \rightarrow \infty$) can itself be compressible at the macroscale (finite $\kappa_d$, $\alpha \rightarrow 1$), because the skeleton can rearrange grains while expelling/drawing in melt from its surroundings. This also means that $\kappa_d$ can be less than $\kappa_l$. 

We have shown that compaction dissipates heat most efficiently when $\kappa_d \sim \kappa_s$, such that $\alpha \ll 1$, and $\zeta$ is close to the critical value (Figs. \ref{fig:heating_solid_bulk_viscosity} and \ref{fig:params_vs_alpha}). To our knowledge, only $\zeta$ has been inferred for Earth-like mantle material \cite{renner2003MeltExtraction, katz2022physics}. Biot's coefficient is indeed a function of melt fraction \cite{selvadurai2020InfluencePore}, as assumed in Eq.~\ref{eq:biot_coeff_phi}, but it is not clear whether $\alpha$ in Io's mantle should approach unity at the critical melt fraction ($b=1$), or some small value ($b=0.01$). If the former is true, meaning that the solid skeleton is easily (de)compacted when the melt fraction is high, then the possibility of significant compaction dissipation is precluded, as small enough $\alpha$ can never be achieved close to the critical compaction viscosity (Fig. \ref{fig:params_vs_alpha}c). How $\alpha$ actually depends on melt fraction is complex, and can be influenced by several factors including the pore geometry \cite{selvadurai2020InfluencePore}. Given the role of Biot's coefficient in controlling compaction heating, the distinction between the single-phase approach of ref.~\cite{kervazo2021SolidTides} and our two-phase approach can be critical. 

If Io's melt is highly incompressible ($\kappa_l > \SI{100}{\giga\pascal}$), then high Darcy and compaction dissipation may be incompatible. We see this by comparing Figs.~\ref{fig:heating_mobility} and \ref{fig:heating_solid_bulk_viscosity} for $\alpha=0.01$ (dashed lines) and $\kappa_l = \SI{200}{\giga\pascal}$ (pink). When the mobility is high, $M_\porosity = \SI{e-3}{\square\metre\per\pascal\per\second}$, Darcy dissipation never exceeds \SI{1}{\giga\watt}, while compaction can reach almost \SI{1}{T\watt} at the critical viscosity. At these high mobilities, Darcy dissipation requires a weak asthenosphere to produce high heating rates, whereas compaction dissipation always requires a strong asthenosphere.

\subsection{Model limitations and future avenues} \label{ssec:diss_limitations_future}

As shown in Figure \ref{fig:patterns}, tidal deformation naturally produces a spatially varying heat distribution. Hence, we expect melt generation to also vary in space, and consequently, melt fraction-dependent parameters such as permeability, shear and compaction viscosity, will too. However, the present formalism assumes that rheological parameters are laterally uniform and depend only on radius, as is typical in tidal deformation studies \cite{sabadini2016GlobalDynamics}. Sophisticated models of tidal deformation that take into account lateral variation in rheological parameters are emerging \cite{beuthe2019EnceladusCrust,rovira-navarro2024SpectralMethod, berne2023InferringMean}, though these do not yet take into account two-phase flow. While this is likely to change the exact spatial distribution of tidal heat generation, it is unlikely to change the overall picture of the importance of Darcy and compaction dissipation, relative to shear heating, in our results.  

As with any tidal heating model, our results are generally sensitive to the rheological model assumed. We explore Io's two-phase tidal deformation under a compacting poro-viscoelastic Maxwell model, but, as discussed above, there may be better choices for a partially molten layer forced at tidal frequencies, such as the Andrade or Sunberg-Cooper models \cite{mccarthy2016TidalDissipation, renaud2017IncreasedTidal}. We caution at choosing a more sophisticated rheology, however, because there is a lack of laboratory attenuation experiments that consider isotropic strain of the material. There is clearly a need to fill this gap if we are to understand the intricacies of Io's tidal heat generation.

Finally, a common assumption that is also made in this work is that the mantle melt fraction is constant in radius and does not vary over a deformation cycle. Neither of these points is likely to be true. Models of radial melt segregation that include a density contrast between solid and melt \cite{spencer2020MagmaticIntrusions, spencer2021TidalControls, miyazaki2022SubsurfaceMagma} all predict the formation of a compaction boundary layer, where the buoyant melt pools at the base of the lithosphere until it is extracted to the surface. This creates a highly non-uniform radial variation in melt fraction across the asthenosphere. Ref.~\cite{kamata2023PoroviscoelasticGravitational} showed that additional Darcy dissipation may arise from these radial porosity gradients (see their Eq.~165). However, as we have mentioned, their treatment of base-state porosity gradients is incomplete as $\porosity$ is assumed constant in their (and our) storage equation \eqref{eq:gov_storage_elastic}. It is thus unclear whether radial porosity gradients will enhance or suppress Darcy and/or compaction dissipation. Given how extreme these gradients may become in a compacting boundary layer, this will be a topic of future work.

\section{Conclusions}\label{sec:conclusion}

In this manuscript, we have combined viscoelastic tidal and poro-viscous compaction theories to develop a tidal deformation model that captures the two-phase dynamics of partially molten planetary interiors. We derive the model, building upon the work of \cite{rovira-navarro2022TidesEnceladus, kamata2023PoroviscoelasticGravitational}, using mathematically consistent constitutive laws that are inspired by the original work of A. E. Love \cite{love1911ProblemsGeodynamics}. 
The model can predict tidal heating from deformation due to shear, melt segregation/Darcy flow, and (de)compaction. We apply the model to the Jovian moon, Io, where we consider Io's asthenosphere to be partially molten with impermeable top and bottom boundaries. We investigate whether compaction and Darcy dissipation within the asthenosphere can contribute substantially to Io's total heat budget.

We find that Darcy dissipation is maximised when the asthenosphere is highly permeable, and the melt is low viscosity. To obtain Darcy heating rates that are comparable to shear heating requires a high enough permeability, which can only be achieved when the asthenosphere's grain size is at least \SI{10}{\cm}, the melt's viscosity is $\SI{1}{\pascal\second}$, and the asthenosphere's melt fraction is between 5--20\%. It is questionable whether such large grain sizes can be reached.  If the melt is highly incompressible, then high Darcy dissipation additionally requires the asthenosphere to be weak to isotropic stresses. Compaction dissipation---heating due to the rearrangement of solid grains---is maximised when the asthenosphere is elastically resistant to isotropic stresses, when compared to the solid grains themselves. This means that high Darcy dissipation and high compaction dissipation may be incompatible, depending on the compressibility of the melt. While the heating patterns generated by Darcy and compaction dissipation are consistent with some interpretations of Io's distribution of volcanic hotspots, we find that compaction and Darcy dissipation are unlikely to account for more than $\sim$1\% of Io's observed total heat output if we assume realistic melt fractions, permeabilities, and liquid viscosities. They may, however, still play a role in Io's convective dynamics. Future work on the evolution of grain size in tidally heated worlds would help further establish the role of Darcy dissipation in these types of planetary body.    

These conclusions are reached by assuming that Io's asthenosphere behaves as a compacting poro-viscoelastic Maxwell material. While it is tempting to apply 
the more experimentally driven Andrade rheology to our model, the lack of laboratory measurements for partially molten rock under isotropic stresses make this approach unsatisfactory at present. We highlight the need for future experiments to measure the attenuation behaviour and poro-viscoelastic parameters of partially molten rock at (or near) tidal frequencies, with both shear and isotropic stresses. 

By including viscous compaction of Io's two-phase asthenosphere, this study makes a first step at capturing the poro-viscoelastic dynamics of a partially molten tidally heated planetary body. This work thus moves us closer to self-consistently modelling the feedback associated with melt in Io's interior, and the inevitable viscous compaction of the mantle as a result. While we find it unlikely that Darcy dissipation can be a major contributor in Io's overall heat budget, melt segregation may still play a role in modulating eruption timing \cite{dekleer2019VariabilityIo}. A future mission to Io would provide an invaluable means towards revealing the details of Io's melt distribution and dynamics \cite{keane2021ScienceCase}. 

\enlargethispage{20pt}

\textbf{Acknowledgements}
We are grateful to A.~Veenstra for discussion and for pointing out a typo in ref.~\cite{kervazo2021SolidTides} for the value of $\phi^*$. This work was supported by a Leverhulme Trust Research Project Grant (RPG-2021-199). For the purpose of open access, the author has applied a CC BY public copyright licence to any author accepted manuscript arising from this submission.


\vskip2pc
\printbibliography

\newpage

\newrefsection

\setcounter{section}{0}
\setcounter{figure}{0}
\setcounter{table}{0}
\counterwithin*{equation}{section}

\newcommand{\snum}{S}

\renewcommand{\theequation}{\snum\arabic{section}.\arabic{equation}}
\renewcommand{\thefigure}{\snum\arabic{figure}}
\renewcommand{\thetable}{\snum\arabic{table}}
\renewcommand{\thesection}{\snum\arabic{section}}

\renewcommand*{\theHsection}{\snum\arabic{section}}
\renewcommand*{\theHfigure}{\snum\arabic{figure}}
\renewcommand*{\theHtable}{\snum\arabic{table}}

\section*{Supplementary Material for \textit{Poro-viscoelastic Tidal Heating of Io}}

This supplementary information contains the details required to reproduce the linearised equations and results in the main text. For convenience, a list of all mathematical symbols is given in Table \ref{tb:params}, and where they are defined in the main text. All equation, section, and table numbers in this document are prefixed with S.

The advective form of the solid constitutive law is derived in \S\ref{app:adv_maxwell}. Support for our assumption of time-invariant porosity is given in \S\ref{app:porosity_assumption}, and in \S\ref{app:storage} we walk through the derivation of the melt storage equation using the solid constitutive law. The tidal-forcing potential arising from orbital eccentricity on a synchronously rotating body is given in \S\ref{app:forcing}. Reduction of the linearised equations into a set of eight coupled ordinary differential equations (ODEs) is summarised in \S\ref{app:ODE}. The numerical implementation of the propagator matrix method to solve these ODEs is described in \S\ref{app:solve_ODEs}. Finally, in \S\ref{app:checks}, we show the difference between the macro and micro approaches to calculating tidal dissipation in the poro-viscoelastic theory from \citet{rovira-navarro2022TidesEnceladus}.
\begin{table}[h!]
    \centering
    \begin{tabular}{lccll}
        \hline
          Quantity & Symbol & Definition & Preferred Value & Units \\ \hline\hline
          Material phase  &  $j$ & $s,\,l$ &   &  -  \\
          Porosity/melt fraction & $\porosity$ &  & 0.1 & - \\
          Phase density & $\rho_j$    &  & $3300$  &  \si{\kg\per\cubic\metre} \\
          Phase-averaged density & $\rho$    & $(1-\phi) \rho_s + \phi \rho_l $  & $3300$  &  \si{\kg\per\cubic\metre} \\
          Velocity & $\vec{v}_j$ & &     & \si{\metre\per\second} \\
          Displacement & $\vec{u}_j$ & &     & \si{\metre} \\
          Segregation flux & $\vec{q}$ & $\porosity (\vec{v}_l - \vec{v}_s)$ &  & \si{\metre\per\second} \\
          Stress   & $\stressT_j $ &  &   & \si{\pascal} \\
          Phase-averaged stress & $\stressT $ & $(1-\phi) \stressT_s + \phi \stressT_l $ &   & \si{\pascal} \\
          Mean stress   & $\sigma_j$  & $\trace{\stressT_j}/3 $ &   & \si{\pascal} \\
          Phase-averaged mean stress & $\sigma $ & $(1-\phi) \sigma_s + \phi \sigma_l $ &   & \si{\pascal} \\
          Pore pressure & $p_l$     &   $-\sigma_l$ &   & \si{\pascal} \\
          Strain        & $\strainT_j$ & Eq. \ref{eq:def_strain} &  & - \\
          Gravitational potential & $\Phi$ &     &   &  \si{\metre\squared\per\second\squared} \\
          Gravity & $g$ &  Eq. \ref{eq:perturb_pot_0}   &   &  \si{\metre\per\second\squared} \\
      \hline 
          \textbf{Rheological parameters} & & & & \\
      \hline 
           Shear modulus   & $\mu_j$ & &  $s,\, l = \left\{ 60, 0\right\}$ & \si{\giga\pascal} \\
           Shear viscosity & $\eta_j$ & &  $s,\, l = \left\{ \num{e21}, 1\right\}$ & \si{\pascal\second} \\
           Bulk modulus & $\kappa_j$ & &  $s,\, l = \left\{ \num{200}, 1\text{--}200\right\}$ & \si{\giga\pascal} \\
           Bulk viscosity & $\zeta$  & & & \si{\pascal\second} \\  
           Biot's coefficient & $\alpha$ & & $0$--$1$ & - \\
           Drained bulk modulus & $\kappa_d$ & $(1-\alpha)\kappa_s$ &  & \si{\giga\pascal} \\
           Storativity & S & Eq. \ref{eq:fourier_storativity_def} & & \si{\per\pascal} \\
           Permeability & $k$ & & & \si{\metre\squared} \\
           Mobility &  $M$  & $k/\eta_l$ & & - \\
           Forcing frequency & $\omega$ &   &  &  \si{rad\per\second} \\
           Grain size/radius & $a$ & & 0.1 & \si{\cm} \\
           Compaction Maxwell time & $\tau_C$ & $\zeta / \kappa_d$ & & \si{\second} \\
      \hline
           \textbf{Constants} & & & & \\
      \hline
              Orbital eccentricity  &  $e$ &   &  0.0048 & - \\
              Orbital period  &  $P$ &   &  42 & \si{hours} \\
              Orbital frequency           &  $\omega$ & $2\pi / P$ & \num{4.1e-5} & \si{rad\per\second} \\
              Rotation frequency           &  $\Omega$ & $\omega$ & \num{4.1e-5} & \si{rad\per\second} \\
              Gravitational constant & $G$  &     &  \num{6.67e-11}  & \si{\metre\cubed\per\kg\per\second\squared} \\
      \hline
    \end{tabular}
    \caption{Main parameter names, symbols, and definitions. }
    \label{tb:params}
\end{table}

\section{Advective Constitutive Laws}\label{app:adv_maxwell}

A general inconsistency in tidal deformation problems is the special treatment of perturbing stresses. Often, perturbations are written so that advection of the hydrostatic stress field is explicitly taken in to account, i.e., $\stressT = \stressT_0 - \dis\cdot\Grad{\stressT_0} + \stressT_1$. Such explicit treatment of advection is not required when perturbing the density field because material advection is naturally included in the continuity equation (Eq. \ref{eq:gov_mass_s}). We show that one approach to this discrepancy is for the constitutive law to also naturally account for advection of hydrostatic stresses. In this section we derive such a constitutive law. As in the main text, we assume perturbations of the form in Eq. \ref{eq:perturb_form}.

We assume that the solid-liquid mixture behaves as a viscously compacting poroviscoelastic Maxwell material. The total strain of the solid in this mixture is given by the sum of its elastic and viscous strains,
\begin{equation}
	\strainT_s = \strainT_{el} + \strainT_{vi}. \label{eq:strain_sum}
\end{equation}
The effective stress is the part of the total stress tensor in the two-phase aggregate that is effective at causing deformation,
\begin{equation} \label{eq:eff_stress_def}
	\stressT' \equiv \stressT + \alpha p_l \identity.
\end{equation}
\noindent The poroviscous rheology states that any deviations from the initial (hydrostatic) effective stress are,
\begin{equation}\label{eq:gov_porovisc}
	\stressT'(\pos) - \stressT'_0 (\pos_0)  = \brackets{\zeta - \frac{2}{3}\eta}\trace{\dot{\strainT}_{vi}}\identity + 2 \eta \dot{\strainT}_{vi}.
\end{equation}
%
The deviation from hydrostatic effective stress at the present position of an element in the deformed state, $\pos$, is relative to that of the material element in the \textit{undeformed} state, $\pos_0$. This is because a material element that undergoes deformation retains its initial hydrostatic pressure \citep{love1911ProblemsGeodynamics}. In the limit where there is no deformation ($\strainT_{vi} = \bm{0}$, $\pos= \pos_0$), the effective stress simply becomes the hydrostatic effective stress in the initial state. 

For a poroelastic rheology, the hydrostatic effective stress deviation is,
\begin{equation} \label{eq:gov_poroelastic}
	\stressT'(\pos) - \stressT'_0 (\pos_0) =  \brackets{\kappa_d - \frac{2}{3}\mu}\trace{\strainT_{el}}\identity + 2 \mu \strainT_{el},
\end{equation}
\noindent where again the stress deviation is relative to an element's hydrostatic effective stress at its position in the undeformed state.  We can expand the left-hand sides of both equations \ref{eq:gov_porovisc} and \ref{eq:gov_poroelastic} in terms of the stresses of the solid and liquid phases. As mentioned in Section \ref{sec:theory}, we explicitly set $\alpha=1$ for viscous stresses as is typical in mantle dynamics \citep{mckenzie1984GenerationCompaction,  katz2022DynamicsPartially}. Thus,
\begin{subequations}
	\begin{align}
		\stressT'(\pos) - \stressT'_0 (\pos_0) &= (1-\phi)\left[\stressT_s(\pos) - \stressT_{s,0}(\pos_{s,0}) \right] + 	(\alpha-\phi)\left[p_l(\pos) - p_{l,0}(\pos_{l,0})\right]\identity, \,\,\, \text{(Elastic)} \\
		\stressT'(\pos) - \stressT'_0 (\pos_0) &= (1-\phi)\Bigl(\stressT_s(\pos) - \stressT_{s,0}(\pos_{s,0}) + 	\left[p_l(\pos) - p_{l,0}(\pos_{l,0})\right]\identity\Bigr), \,\,\, \text{(Viscous)}
	\end{align}
\end{subequations}
where we crucially recognise that the initial position of the material phases within an element in the deformed state are phase dependent.
Using Equation \ref{eq:pos_vec_def} we can expand the hydrostatic stresses with a Taylor series expansion so that the stress perturbations become,
\begin{subequations}\label{eq:stress_pert_twophase}
	\begin{align}
		\stressT'(\pos) - \stressT'_0 (\pos_0) &\approx (1-\phi)\Bigl(\stressT_s(\pos) - \stressT_{s,0}(\pos) + \disS\cdot\Grad{\stressT_{s,0}}\Bigr) + 	(\alpha-\phi)\Bigl(p_l(\pos) - p_{l,0}(\pos) + \disL \cdot \Grad{p}_{l,0}\Bigr)\identity, \,\,\, \text{(Elastic)} \\
		\stressT'(\pos) - \stressT'_0 (\pos_0) &\approx (1-\phi)\Bigl(\stressT_s(\pos) - \stressT_{s,0}(\pos) + \disS\cdot\Grad{\stressT_{s,0}} + 	\left[p_l(\pos) - p_{l,0}(\pos) + \disL \cdot \Grad{p}_{l,0}\right]\identity\Bigr). \,\,\, \text{(Viscous)}
	\end{align}
\end{subequations}
The poroelastic and poroviscous constitutive laws then become,
\begin{subequations}
	\begin{align}
		(1-\phi)&\Bigl(\stressT_s(\pos) - \stressT_{s,0}(\pos) + \disS\cdot\Grad{\stressT_{s,0}}\Bigr) + 	(\alpha-\phi)\Bigl(p_l(\pos) - p_{l,0}(\pos) + \disL \cdot \Grad{p}_{l,0}\Bigr)\identity \notag \\
		&= \brackets{\kappa_d - \frac{2}{3}\mu}\trace{\strainT_{el}}\identity + 2 \mu \strainT_{el}, \,\,\, \text{(Elastic)}  \label{eq:elastic_adv} \\
		(1-\phi)&\Bigl(\stressT_s(\pos) - \stressT_{s,0}(\pos) + \disS\cdot\Grad{\stressT_{s,0}} + \left[p_l(\pos) - p_{l,0}(\pos) + \disL \cdot \Grad{p}_{l,0}\right]\identity\Bigr)\notag \\ 
		&= \brackets{\zeta - \frac{2}{3}\eta}\trace{\dot{\strainT}_{vi}}\identity + 2 \eta \dot{\strainT}_{vi}. \,\,\, \text{(Viscous)} \label{eq:viscous_adv}
	\end{align}
\end{subequations}
\noindent Equation \ref{eq:elastic_adv} reduces to Eq. 6 in Chapter VII of \citet{love1911ProblemsGeodynamics} when $\phi = \alpha = 0$ and $\kappa_d = \kappa$. 

Taking the time derivative of \eqref{eq:strain_sum} and \eqref{eq:elastic_adv}, and with some algebra, we can form the poroviscoelastic constitutive law,
\begin{align}
	&(1 - \phi) \left[ \dot{\stressT}_s + \vec{v}_s \cdot \Grad{\stressT_{s,0}}\right] + (\alpha - \phi)\left[\dot{p}_{l} + \vec{v}_l \cdot \Grad{p_{l,0}} \right]\identity + (1 - \phi)\frac{\mu}{\eta} \left[\stressT_{s} - \frac{1}{3} \trace{\stressT_s}\identity\right] \notag \\ 
	&+ (1 - \phi)\frac{\kappa}{\zeta} \left[\frac{1}{3} \trace{\stressT_s - \stressT_{s,0} + \dis_s \cdot \Grad{ \stressT_{s,0}} } + (p_l - p_{l,0} + \dis_l \cdot \Grad{p_{l,0}})\right]\identity = \lambda_d \trace{\dot{\strainT}_s} \identity + 2 \mu \dot{\strainT}_s .
\end{align}

\noindent Noting that $\dot{\stressT}_j = \dot{\stressT}_{j,1}$ and $D_j \stressT_j / Dt \approx \dot{\stressT}_{j,1} + \vec{v}_j \cdot \Grad{\stressT_{j,0}}$ when the stress perturbation and displacements are small, then the above constitutive law can be rewritten as,
\begin{align}
	(1 - \phi) \frac{D_s \stressT_s}{D t} &+ (\alpha - \phi)\frac{D_l p_{l}}{Dt} \identity + (1 - \phi)\frac{\mu}{\eta} \left[\stressT_{s} - \frac{1}{3} \trace{\stressT_s}\identity\right] \notag \\ 
	&+ (1 - \phi)\frac{\kappa}{\zeta} \left[\frac{1}{3} \trace{\stressT_s - \stressT_{s,0} + \dis_s \cdot \Grad{ \stressT_{s,0}} } + (p_l - p_{l,0} + \dis_l \cdot \Grad{p_{l,0}})\right]\identity = \lambda_d \trace{\dot{\strainT}_s} \identity + 2 \mu \dot{\strainT}_s.
\end{align}
This constitutive law is desirable because (1) it naturally accounts for advection of the initial solid and liquid hydrostatic stresses by deformation, and (2) if the stress is the hydrostatic base state everywhere (i.e., $\stressT_{j} = \stressT_{j,0}$), then no deformation occurs and our base state is consistent with the assumption that $\dis_j = \vec{0}$. We can write the compacting parts of the constitutive law in integral form to give a slightly cleaner expression,
\begin{align}
	(1 - \phi) &\left[\frac{D_s \stressT_s}{D t} + \frac{1}{3}\frac{\kappa_d}{\zeta} \int_t \frac{D_s \trace{\stressT_s}}{D t}dt \identity\right] + \left[(\alpha - \phi)\frac{D_l p_{l}}{Dt} + (1 - \phi)\frac{\kappa_d}{\zeta} \int_t \frac{D_l p_l}{D t}dt \right] \identity \notag \\
	&+ (1 - \phi)\frac{\mu}{\eta} \left[\stressT_{s} - \frac{1}{3} \trace{\stressT_s}\identity\right] = \lambda_d \trace{\dot{\strainT}_s} \identity + 2 \mu \dot{\strainT}_s.
\end{align}
If we substitute $\stressT_j = \stressT_{j,0} + \stressT_{j,1}$ and linearise we obtain the constitutive law of the stress perturbation,
\begin{align}
	(1 - \phi)\left[ \dot{\stressT}_{s,1} + \vec{v}_s \cdot \Grad{\stressT_{s,0}} \right]&+ (\alpha - \phi)\left[\dot{p}_{l,1} + \vec{v}_l \cdot \Grad{p_{l,0}}\right] \identity + (1 - \phi)\frac{\mu}{\eta} \left[\stressT_{s,1} - \frac{1}{3} \trace{\stressT_{s,1}}\identity\right] \notag \\ 
	&+ (1 - \phi)\frac{\kappa}{\zeta} \left[\frac{1}{3} \trace{\stressT_{s,1} + \dis_s \cdot \Grad{ \stressT_{s,0}} } +  (p_{l,1} + \dis_l \cdot \Grad{p_{l,0}})\right]\identity = \lambda_d \trace{\dot{\strainT}_s} \identity + 2 \mu \dot{\strainT}_s .
\end{align}
Assuming that the base state stresses are hydrostatic, we recover Equation \ref{eq:perturb_const_s_2}.

\section{Assumption of constant porosity in time}\label{app:porosity_assumption}

In this work assume that porosity is constant in time, such that as the mantle is deformed, changes in melt fraction are negligible (i.e., $\porosity_1 = 0$). A more rigorous treatment would specifically include this variation by using $\porosity = \porosity_0 + \porosity_1$ in the linearisation of the governing equations. The liquid-phase continuity equation \eqref{eq:gov_mass_l} can be combined with an equation of state \eqref{eq:eos_liquid} to yield an equation for the evolution of melt fraction,
\begin{equation}
	\rho_l \frac{\partial \porosity}{\partial t} + \porosity \frac{\rho_l}{\kappa_l} \frac{\partial p_l}{\partial t} = -\Divb{\porosity \rho_l \vec{v}_l}.
\end{equation}
Linearising this expression and assuming incompressible liquid yields a simplified evolution equation for the porosity perturbation. To order of magnitude, the porosity perturbation is then $\porosity_1 \sim 2\pi \porosity_0 \vec{v}_l / \omega R$. For Io-like parameters, a background melt fraction of $\phi_0 \sim 0.1$, and a generous liquid velocity of $\SI{1}{\mm\per\second}$, gives $\phi_1 \sim \num{e-5}$. This is sufficiently small that it can be safely neglected, justifying our assumption to keep $\porosity$ constant.

\section{Deriving the Liquid Constitutive Law} \label{app:storage}

We note that material derivatives between liquid and solid can be related using the relative velocity between phases,
\begin{equation}
	\frac{D_l}{Dt} = \frac{D_s}{Dt} + \vec{v}_{rel} \cdot \Grad{} \, .
\end{equation}
This allows us to express the solid constitutive law in \eqref{eq:gov_maxwell_compaction} in terms of the total stress,
\begin{align}
	&\left[\frac{D_s \stressT + \alpha p_l \identity}{D t} + \frac{1}{3}\frac{\kappa_d}{\zeta} \int_t \frac{D_s \trace{\stressT} + 3 p_l}{D t}dt \identity\right] + \left[(\alpha - \phi)\vec{v}_{rel} \cdot \Grad{p_l} + (1 - \phi)\frac{\kappa_d}{\zeta} \int_t \vec{v}_{rel} \cdot \Grad{p_l} dt\right] \identity \notag \\
	&+ (1 - \phi)\frac{\mu}{\eta} \left[\stressT_{s} - \frac{1}{3} \trace{\stressT_s}\identity\right] = \lambda_d \trace{\dot{\strainT}_s} \identity + 2 \mu \dot{\strainT}_s.
\end{align}


\noindent Taking the trace of this expression and rearranging gives the evolution of mean effective stress,
\begin{align}
	\frac{D_s \sigma}{D t}    
	&  = - \alpha\frac{D_s  p_l }{D t} - \frac{\kappa_d}{\zeta} \int_t \frac{D_s \sigma + p_l}{D t}dt \notag \\
	& -  \left[(\alpha - \phi)\vec{v}_{rel} \cdot \Grad{p_l} + (1 - \phi)\frac{\kappa_d}{\zeta} \int_t \vec{v}_{rel} \cdot \Grad{p_l} dt\right] +  \kappa_d \trace{\dot{\strainT}_s} .
\end{align}


Substituting $\sigma = \sigma_0 + \sigma_1$, $p_l = p_{l,0} + p_{l,1}$, and neglecting second order terms in perturbed quantities, and recognising that base-state quantities are independent of time gives,

\begin{align}
	\dot{\sigma}_1 + \vec{v}_s \cdot \Grad{\sigma_0}    
	&\approx - \alpha\left(\dot{p}_{l,1} + \vec{v}_s \cdot \Grad{p_{l,0}} \right) - \frac{\kappa_d}{\zeta} \biggl[ \sigma_1 + p_{l,1} + \vec{u}_s \cdot \Gradb{\sigma_0 + p_{l,0}}\biggr] \notag \\
	& -  (\alpha - \phi)\vec{v}_{rel} \cdot \Grad{p_{l,0}} - (1 - \phi)\frac{\kappa_d}{\zeta} \biggl[ \vec{u}_{rel} \cdot \Grad{p_{l,0}} \biggr] +  \kappa_d \trace{\dot{\strainT}_s} 
\end{align}
The Fourier transform of this, after some rearrangement and assuming base state stresses are hydrostatic and depend only on radius, gives Equation \ref{eq:fourier_maxwell_trace_total} which can be written as,
\begin{equation}\label{eq:fourier_maxwell_trace2}
	\f{\sigma}_1 + \f{u}_s^r \rho_0 g = - \f{\alpha} \f{p}_{l,1} + \brackets{\f{\alpha} \f{u}_l^r - \phi u_{rel}^r} \rho_{l,0} g + \f{\kappa}_d \trace{\f{\strainT}_s}
\end{equation}
The Fourier transform of the pore pressure perturbation equation \eqref{eq:perturb_storage} is,
\begin{equation}\label{eq:fourier_storage2}
	\left[\frac{\phi}{\kappa_l} - \frac{\phi}{\kappa_s} \right] \left[ \f{p}_{l,1} - \f{u}_l^r \rho_{l,0} g \right] =  \frac{\phi}{\kappa_s} \f{u}^r_{rel} \rho_{l,0} g  +  \frac{1}{\kappa_s} \left[\f{\sigma}_1 + \rho_0 \f{u}_s^r g \right] -  \phi \trace{\f{\strainT}_{rel}}  -  \trace{\f{\strainT}_{s}}.
\end{equation}
If we substitute Eq. \ref{eq:fourier_maxwell_trace2} into Eq. \ref{eq:fourier_storage2} and rearrange for $\f{p}_{l,1}$, we arrive at the complete storage equation with compaction, Equation \ref{eq:fourier_storage}.



\section{Coupled System of ODEs}\label{app:ODE}

Following the classic approach (e.g., \cite{takeuchi1972SeismicSurface}), we expand the unknown parameters in terms of scalar and vector spherical harmonics of degree $n$ and order $m$. That is,
\begin{subequations}\label{eq:sph_harm_exp_def}
	\begin{align}
		\f{\dis}_s &= \sum_{n,m}\s{\pot}^T_{nm} \left(\s{u}^r_{s,nm}(r) \Rnm + \s{u}^t_{s,nm}(r) \Snm \right), \\
		\f{\stressT}_1^\delta\cdot \rhat &= \sum_{n,m}\s{\pot}^T_{nm} \left(\s{\sigma}^{rr}_{nm}(r) \Rnm + \s{\sigma}^{rt}_{s,nm}(r) \Snm \right), \\
		\f{\pot} &= \sum_{n,m}\s{\pot}^T_{nm} \s{\pot}_{nm}(r) \Ynm ,  \label{eq:sph_harm_pot}  \\
		\f{\dis}_{\text{rel}} &= \sum_{n,m}\s{\pot}^T_{nm}  \left(\s{u}^r_{\text{rel},nm}(r) \Rnm + \s{u}^t_{\text{rel},nm}(r) \Snm \right), \\
		\f{p}_{l,1}^\delta &= \sum_{n,m}\s{\pot}^T_{nm} \s{p}_{l,nm}(r) \Ynm  ,
	\end{align}
\end{subequations}
\noindent where the unnormalised complex spherical harmonic $\Ynm = P_{nm}(\cos\clat) e^{im\lon}$, $P_{nm}$ is the unnormalised Associated Legendre function, and the spherical harmonic vectors are,
\begin{align}
	\Rnm &= \Ynm \rhat, \\
	\Snm &= r \Grad{}_2 \Ynm, 
\end{align}
\noindent where $\Grad{}_2$ is the gradient operator for only the $\thetahat$ and $\phihat$ components. In Equation \ref{eq:sph_harm_exp_def}, quantities with $\hat{}$ symbols are the degree $n$ and order $m$ spherical harmonic expansion coefficients of the equivalent Fourier transformed variable. The superscripts $r$ and $t$ denote the radial and tangential components of the vector spherical harmonic expansion, respectively. Following \cite{kamata2023PoroviscoelasticGravitational}, we set $\s{\pot}_{nm}^T$ as the amplitude of the tidal forcing for a particular $n$ and $m$. 

With the above expansion, the problem can be reduced to a coupled set of 8 ODEs, as shown for the poroviscoelastic problem by \RNporo and \Kporo. Deriving these ODE's is clearly outlined in \Kporo, and we do not repeat that derivation here. We provide the set of ODEs solved in this work below, and highlight our differences from \Kporo and \RNporo. 

As standard \citep[e.g][]{takeuchi1972SeismicSurface}, we cast the problem's unknowns \eqref{eq:sph_harm_exp_def} in a solution vector, 
\begin{align}\label{eq:y_def}
	\nm{\vec{y}} &= \brackets{\s{u}^r_{s,nm} ,\,
		\s{u}^t_{s,nm},\, 
		\s{\sigma}^{rr}_{nm},\, 
		\s{\sigma}^{rt}_{nm},\, 
		-\s{\Phi}_{nm},\, 
		Q_{nm},\, 
		\s{p}_{l,nm},\, 
		\s{u}^r_{rel,nm}   }^t \notag \\
	&= \brackets{y_1,\, y_2,\, y_3,\, y_4,\, y_5,\, y_6,\, y_7,\, y_8}^t.
\end{align}
The tangential segregation velocity, $y_9 = \s{u}^t_{rel,nm}$, can be expressed entirely in terms of the other unknowns \eqref{eq:y_system9}, hence why it need not explicitly appear in the solution vector. The ``potential stress'' is defined as \cite{sabadini2016GlobalDynamics, rovira-navarro2022TidesEnceladus,kamata2023PoroviscoelasticGravitational},
\begin{equation}
	Q_{nm} = y_6 = \Gradr{y_5} + \frac{n+1}{r}y_5 - 4\pi G (\rho_0 y_1 + \porosity \rho_{l,0} y_8).
\end{equation}
In terms of notation and approach, there are three important differences from both \cite{kamata2023PoroviscoelasticGravitational} and \cite{rovira-navarro2022TidesEnceladus}; (1) like \cite{kamata2023PoroviscoelasticGravitational}, we form the problem around fluid displacements, rather than velocities as used in \cite{rovira-navarro2022TidesEnceladus} as this makes the resulting algebraic system better conditioned; (2) our $y_2$ denotes tangential displacement and $y_3$ denotes radial stress following \cite{rovira-navarro2022TidesEnceladus}, but opposite to \cite{kamata2023PoroviscoelasticGravitational}; and (3) our $y_7$ corresponds to pore pressure and $y_8$ to relative radial displacement, which is consistent with \cite{rovira-navarro2022TidesEnceladus} and opposite to \cite{kamata2023PoroviscoelasticGravitational}. 

Substituting Eq. \ref{eq:y_def} into the linearised and Fourier transformed equations, and working through substantial algebra yields,

\begin{subequations}\label{eq:y_system1_8}
	\begin{align}
		\frac{dy_1}{dr} &= - \frac{2\f{\lambda}_d}{\f{\beta}r} y_1 +  \frac{n(n+1)\f{\lambda}_d}{\f{\beta}r} y_2 + \frac{1}{\f{\beta}} y_3 + \frac{\f{\alpha}}{\f{\beta}} y_7 \\
		\frac{dy_2}{dr} &= - \frac{1}{r} y_1 + \frac{1}{r}y_2 + \frac{1}{\f{\mu}}y_4 \\
		\frac{dy_3}{dr} 
		&= 
		\left[
		\frac{12\f{\kappa}_d \f{\mu}}{(\lambda_d + 2\mu)r^2} - 4 \frac{\rho_{0} g}{r} + i\frac{k \rho^2_{l,0} g^2 n(n+1)}{ \omega  \eta_l r^2} \right] y_1 
		+  \frac{ n(n+1)}{r} \left[\rho_{0} g - \frac{\f{\mu}}{r}\frac{6\f{\kappa}_d}{\f{\beta}}  \right] y_2 
		\notag  \\
		& \quad - \frac{4\mu}{\f{\beta}r} y_3 
		+ \left[\frac{n(n+1)}{r}\right]y_4   \notag 
		+ \frac{n+1}{r}\left[\rho_0 - i\frac{k \rho^2_{l,0} g\, n}{ \omega  \eta_l r}\right] y_5  
		- \rho_0 y_6 
		\notag  \\
		& \quad    
		+ \left[i\frac{k \rho_{l,0} g\, n(n+1)}{ \omega  \eta_l r^2} -\frac{4 \f{\mu} \f{\alpha}}{\f{\beta}r}\right] y_7
		+ \left[i\frac{k \rho^2_{l,0} g^2 n(n+1) }{ \omega  \eta_l r^2} - 4\phi   \frac{\rho_{l,0} g}{r}\right]y_8   \\ 
		\frac{dy_4}{dr} 
		&=  
		\left[ \frac{\rho_{0} g}{r} -\frac{6\f{\mu}\f{\kappa}_d}{\f{\beta}r^2} \right] y_1 
		+  \frac{2\f{\mu}}{r^2}\left[\frac{2n(n+1)(\f{\lambda}_d + \f{\mu})}{\f{\beta}} - 1 \right] y_2 
		- \frac{\f{\lambda}_d}{\f{\beta}r} y_3 
		- \frac{3 }{r}y_4
		- \frac{\rho_0}{r} y_5\notag \\
		&\qquad  
		+ \frac{2\f{\mu}\f{\alpha}}{\f{\beta}r} y_7
		+ \frac{\phi \rho_{l,0}g}{r}y_8 \\
		\frac{d y_5}{dr} &= 
		4\pi G \rho_0 y_1 -  \frac{n+1}{r}y_5  + y_6 + 4\pi G \porosity \rho_{l,0} y_8  \\
		\frac{dy_6}{dr} 
		&= 
		\frac{4\pi G (n+1)}{r}\left[\den_{0} -  i\frac{n k \rho^2_{l,0} g}{ \omega   \eta_l r}  \right] y_1 
		- 4\pi G \den_{0} \frac{n(n+1)}{r} y_2 
		+ 4\pi G i\frac{k \rho^2_{l,0}n(n+1)}{ \omega   \eta_l r^2}  y_5
		+ \frac{n-1}{r}y_6 \notag \\ 
		& \quad 
		- 4\pi G i\frac{k \rho_{l,0}n(n+1)}{ \omega   \eta_l r^2} y_7 
		+ \frac{4\pi G (n+1)}{r} \left[ \phi\denL{0}  -  i\frac{n k \rho^2_{l,0} g }{ \omega   \eta_l r}  \right]y_8    \\
		\frac{dy_7}{dr} &= 	\frac{\rho_{l,0}g}{r}\left[4 - i\frac{k \rho_{l,0} g n(n+1)}{ \omega \phi  \eta_l r}\right] y_1
		- \left[\frac{n(n+1)}{r} \rho_{l,0} g \right] y_2   
		- \rho_{l,0}\frac{n+1}{r}\left[1  - i\frac{k \rho_{l,0} g n}{ \omega \phi  \eta_l r} \right]y_5  
		+ \rho_{l,0} y_6 
		\notag  \\
		&\quad 
		- i\frac{k \rho_{l,0} g n(n+1)}{ \omega \phi  \eta_l r^2} y_7
		+\left[ - i \omega \frac{\eta_l}{ k } \phi  - 4\pi G(\rho_0 - \phi \rho_{l,0}) \rho_{l,0} + \frac{\rho_{l,0}g}{r}\left[4  -i\frac{k \rho_{l,0} g n(n+1) }{ \omega \phi  \eta_l r}\right]\right] y_8      \\ 
		\frac{dy_8}{dr} 
		&= 
		\frac{1}{r}\left[ i\frac{k \rho_{l,0} g n(n+1)}{ \omega \phi  \eta_l r} -\frac{\f{\alpha}}{\phi}\frac{4\f{\mu}}{\f{\beta}}\right] y_1 
		+ \frac{\f{\alpha}}{\phi}\frac{2n(n+1)\f{\mu}}{\f{\beta}r} y_2 
		- \frac{\f{\alpha}}{\phi}\frac{1}{\f{\beta}} y_3 
		- i\frac{k \rho_{l,0} n(n+1)}{ \omega \phi  \eta_l r^2}  y_5
		\notag \\
		& \quad 
		+\left[i\frac{k n(n+1)}{ \omega \phi  \eta_l r^2} - \frac{1}{\phi}\left[\f{S} + \frac{ \f{\alpha}^2 }{\f{\beta}}\right] \right] y_7 
		+\left[i\frac{k \rho_{l,0} g n(n+1) }{ \omega \phi  \eta_l r^2} -  \frac{2}{r}  \right]y_8   
	\end{align}
\end{subequations}
\noindent where $\f{\beta} = \f{\lambda}_d + 2\f{\mu}$. The above equations are equivalent to \cite{kamata2023PoroviscoelasticGravitational} Eqs. 57, 59, 61, and C1-C5, after swapping their $y_2$ with $y_3$ and their $y_9$ with $y_7$. The relative tangential displacement is given by,
\begin{equation}\label{eq:y_system9}
y_9  = i\frac{k }{ \omega \phi  \eta_l r} \left[\rho_{l,0} g  y_1 + \rho_{l,0} g y_8   
+  y_7     - \rho_{l,0}  y_5 \right]. \notag
\end{equation}
The equations in \ref{eq:y_system1_8} must be radially integrated to obtain the solution, which is outlined in the next section. Once obtained, the potential Tidal Love number can be determined at the surface with \citep{sabadini1982PolarWandering}.,
\begin{equation}\label{eq:k2_from_ode}
k_2 = y_5(R) - 1.
\end{equation}

\section{Numerical Solution} \label{app:solve_ODEs}

The tidal response is calculated by numerically integrating the coupled set of ODE's from the Core--Mantle Boundary (CMB) to the surface. The system of ODE's in \ref{eq:y_system1_8} is written in the compact form
\begin{equation}\label{eq:ode_diff_form}
\frac{d\vec{y}_{nm}}{dr} = \mat{A}_{n}(r) \vec{y}_{nm},
\end{equation}
where the indices of $\mat{A}_n(r)$ are obtained from the equations in \ref{eq:y_system1_8}. To ease the notation, we will now drop the $n$ and $m$ subscripts. There are four integration constants that must be found to enforce the boundary conditions in the problem. Finding these constants depends on the three (single-phase) or four (two-phase) linearly independent solutions to the above equation. We write these linearly independent solutions as
\begin{equation}
\mat{Y}(r) = \begin{pmatrix}
	\vec{y}_1(r) & \vec{y}_2(r) & \vec{y}_3(r) & \vec{y}_4(r)
\end{pmatrix},
\end{equation}
where each column of the $8\times 4$ matrix $\mat{Y}$ represents one linearly independent solution to Equation \ref{eq:ode_diff_form}. The ODE defining the problem can then be written as
\begin{equation}\label{eq:ode_diff_linearly_ind}
\frac{d \mat{Y}(r)}{dr} = \mat{A}(r) \mat{Y}(r),
\end{equation}
which we solve using a standard Runge-Kutta 4th-order (RK4) accurate numerical integration scheme.  

To apply the RK4 method, the interior of Io must be discretised. To some extent, this has already occurred by dividing Io into core, mantle, asthenosphere and crust (Table \ref{tb:io_interior}). We denote these as primary layers with index $i$, where the core is $i=0$ and crust is $i=3$. Material parameters are held constant within each primary layer. To ensure numerical accuracy, each primary layer $i$ is also subdivided into $N_i$ discrete secondary layers (and $N_i+1$ boundaries), which we denote with index $j$. The radial position at the bottom boundary of the $j$th secondary layer within the $i$th primary layer is therefore given by $r_{i,j}$. We set the radial grid spacing $\Delta r_i$ as a constant within each primary layer, such that $r_{i,j} + \Delta r_i = r_{i,j+1}$.

In practice, integrating Equation \ref{eq:ode_diff_linearly_ind} with an RK4 (or other integration) scheme means that the solution at the bottom of the discretised $j$th layer is ``propagated'' to the top of the same layer. This operation can be written as,
\begin{equation}\label{eq:rk4_def}
\mat{Y}(r_{i,j}+\Delta r_i) = \mat{B}(r_{i,j})\mat{Y}(r_{i,j}),
\end{equation}
where, for the RK4 method,
\begin{equation}
\mat{B}(r_{i,j}) \equiv \identity + \frac{1}{6} \left( \mat{K}_1 + \mat{K}_2 + \mat{K}_3 + \mat{K}_4\right)
\end{equation}

\noindent and,
\begin{subequations}
\begin{align}
	\mat{K}_1 &= \Delta r_i \mat{I}_{\phi}(r_{i,j})\mat{A}(r_{i,j})\mat{I}_{\phi}(r_{i,j}) \\
	\mat{K}_2 &= \Delta r_i \mat{I}_{\phi}(r_{i,j})\mat{A}(r_{i,j} + \Delta r_i / 2) \mat{I}_{\phi}(r_{i,j})\left[\identity + \frac{1}{2}\mat{K}_1\right] \\
	\mat{K}_3 &= \Delta r_i \mat{I}_{\phi}(r_{i,j})\mat{A}(r_{i,j} + \Delta r_i / 2) \mat{I}_{\phi}(r_{i,j})\left[\identity + \frac{1}{2}\mat{K}_2\right] \\
	\mat{K}_4 &= \Delta r_i \mat{I}_{\phi}(r_{i,j})\mat{A}(r_{i,j} + \Delta r_i) \mat{I}_{\phi}(r_{i,j})\left[\identity + \mat{K}_3\right].
\end{align}
\end{subequations}
The coefficient matrix within primary layer $i$, $\mat{A}(r_{i,j})$, is calculated using the material parameters of the same layer (e.g., $\mat{A}(r_{i,j}) = f(\f{\mu}_i,\, \rho_i,\, \f{\kappa}_{d,i},\, ...)$ ). The identity-like matrix $\mat{I}_\phi$ has elements,
\begin{equation}
I_{\phi,i',j'} (r_{i,j}) = \begin{cases}
	\delta_{i',j'},& (i'=1,\, 2,\, \dots 6),\\
	\delta_{i',i_\phi}\delta_{i',j'},  & (i' = 7,\, 8),
\end{cases}
\end{equation}
where $\delta_{ij}$ is the Kronecker delta function, and $i_\phi$ is the index of the porous layer. Thus $\mat{I}_\phi$ is simply the identity matrix when in a porous layer, but in a non-porous layer it acts to set to zero the 7th and 8th columns and rows of $\mat{A}$, hence turning off any two-phase calculation. Thus, when $\phi = \alpha = \rho_l = 0$ and $i = i_\phi$, $\mat{I}_\phi\mat{A}\mat{I}_\phi$ gives the $6\times6$ coefficient matrix in the standard solid body problem \citep[e.g.,][]{sabadini2016GlobalDynamics}, with two extra rows and columns of zeros. Numerically, this step is unnecessary and inefficient as the matrix multiplication can be applied over only indices 1:6. 


To continue the integration from one secondary layer to the next, continuity of the solution is enforced,
\begin{equation}
\mat{Y} (r_{i,j+1}) = \mat{Y} (r_{i,j} + \Delta r_i). 
\end{equation}
Applying this recursively from the bottom  to top of the $i$th primary layer ($j=1$ to $j=N_i+1$)  yields,
\begin{equation}\label{eq:ode_Y_propagate}
\mat{Y} (r_{i,N_j+1})= \left(\overset{\curvearrowleft}{\prod_{j=1}^{N_i}}\mat{B}(r_{i,j})\right)\mat{Y}(r_{i,1}).
\end{equation}
where the arrow indicates that additional products are multiplied from the left-hand side.

At a boundary between two primary layers, continuity is also enforced,
\begin{equation}\label{eq:ode_continuity}
\mat{Y} (r_{i+1,1}) = \mat{Y} (r_{i,N_i+1}). 
\end{equation}
If the $i$th layer is solid and the $i+1$th layer has two phases, the continuity condition changes to,
\begin{equation}\label{eq:ode_continuity_porous}
\mat{Y}(r_{i+1,j}) = \mat{Y} (r_{i,N_i+1}) + \mat{B}_p,
\end{equation}
\noindent where the $8\times4$ porous matrix $\mat{B}_p$ has indices $B_{p,i'j'} = \delta_{i'7}\delta_{j'4}$ (i.e., $1$ on row seven, column 4, and zeros elsewhere). This ensures that the pore pressure ($y_7$) is non-zero at the boundary (\RNporo).


%

For the four layer interior structure explored here (Table \ref{tb:io_interior}) with core, lower mantle ($i=1$), partially molten asthenosphere ($i=i_\phi=2$), and crust ($i=3$), the $\mat{Y}$ solution within any layer is obtained by recursively applying Eq. \ref{eq:ode_Y_propagate} along with the correct continuity condition (Eqs. \ref{eq:ode_continuity} or \ref{eq:ode_continuity_porous}).
The linearly independent solutions at position $j$ in each layer are then,
\begin{subequations}\label{eq:ode_integrate_layers}
\begin{align}
	\vec{Y}(r_{C}) &= \mat{Y} (r_{1,1}), & &\text{(CMB)} \label{eq:ode_integrate_layers_cmb} \\
	\vec{Y}(r_{1, j}) &= \mat{D} (r_{1, j}) \mat{Y} (r_{1,1}), & &(\text{lower mantle}, 1\leq j\leq N_1+1) \label{eq:ode_integrate_layers_mantle} \\
	\vec{Y}(r_{2, j}) &= \mat{D}(r_{2, j}) \left[\mat{D}_1 \mat{Y} (r_{1,1}) + \vec{B}_p \right], & &(\text{asthenosphere}, 1\leq j\leq N_2+1) \label{eq:ode_integrate_layers_asthen}\\
	\vec{Y}(r_{3, j}) &= \mat{D} (r_{3, j}) \mat{D}_2 \left[\mat{D}_1 \mat{Y} (r_{1,1}) + \vec{B}_p \right] , & &(\text{crust}, 1\leq j\leq N_3+1) \label{eq:ode_integrate_layers_crust}
\end{align}	
\end{subequations}
\noindent where the propagator matrix is,
\begin{equation}
\mat{D}(r_{i,j}) \equiv 
\begin{cases}
	\identity,& (j = 1),\\
	\overset{\curvearrowleft}{\prod}_{j'=1}^{j-1}\mat{B}(r_{i,j'}),  & (1 < j \leq N_i +1),
\end{cases}
\end{equation}
which guarantees the solution at the lower boundary is retrieved when $j=1$, and, for shorthand, the propagator matrix used to traverse from the bottom to top of the entire $i$th layer is,
\begin{equation}
\mat{D}_i \equiv \mat{D}(r_{i,N_i+1}) = \overset{\curvearrowleft}{\prod_{j'=1}^{N_i}}\mat{B}(r_{i,j'}).
\end{equation}

For an inviscid, liquid core, the unconstrained solution at the CMB is \citep[][Eq. 63]{sabadini1982PolarWandering},
\begin{equation}
\mat{Y}_{r_C} = \begin{pmatrix}
	-r^n / g& 0 & 1 & 0 \\
	0 & 1 & 0 & 0 \\
	0 & 0 & g \rho & 0 \\
	0 & 0 & 0 & 0 \\
	r^n & 0 & 0 & 0 \\
	2(n-1) r^{n-1}& 0  & 4\pi G \rho & 0 \\
	0 & 0 & 0 & 0 \\
	0 & 0 & 0 & 0 
\end{pmatrix}.
\end{equation}

\noindent Integrating Equation \ref{eq:ode_Y_propagate} introduces four constants of integration, $\vec{c} = (c_1, c_2, c_3, c_4)^t$. To determine these constants, four boundary conditions must be applied. The first three boundary conditions are applied at the surface on $y_3$, $y_4$, and $y_6$, while the fourth is applied at the top of the partially molten layer on $y_8$ (\RNporo, \Kporo). We write these boundary conditions in the vector,
\begin{equation}
\vec{b} = \begin{pmatrix}
	b_3(R) \\ b_4(R) \\ b_6(R) \\ b_8(r_\phi)
\end{pmatrix}
= 
\begin{pmatrix}
	0 \\ 0 \\ 2(n+1) / R \\ 0
\end{pmatrix},
\end{equation}
\noindent which corresponds to zero and radial stress conditions at the surface ($b_3 = b_4 = 0$), a tidal forcing condition on the potential stress ($b_6 = 2(n+1)/R$), and zero radial segregation velocity at the impermeable top of the partially molten layer ($b_8 = 0$) (\RNporo, \Kporo).


To apply these boundary conditions, we need to extract the relevant components of $\mat{Y}$ at $R$ and $r_\phi$. The linearly independent segregation fluxes at $r_\phi$ are obtained by first evaluating Eq. \ref{eq:ode_integrate_layers_asthen} at $r_{2,N_2+1} (= r_\phi)$, and then multiplying by a projection matrix,
\begin{equation}\label{eq:num_extract_porous}
\mat{Y}_{8}(r_\phi) = \mat{P}_8 \mat{D}_2 \left[\mat{D}_1 \mat{Y}_{r_C} + \mat{B}_p \right].
\end{equation}
\noindent Here, $\mat{P}_8$ is a $4\times8$ projection matrix with components $P_{8,ij} = \delta_{i4}\delta_{j8}$, which extracts the radial relative displacements solutions from $\mat{Y}(r_\phi)$. The four linearly independent surface solutions for the stresses and potential stress are obtained in similar fashion from Eq. \ref{eq:ode_integrate_layers_crust} at $r_{3,N_3+1} (= R)$,
\begin{equation}\label{eq:num_extract_surf}
\mat{Y}_{3,4,6} (R) = \mat{P} \mat{D}_3 \mat{D}_2 \left[\mat{D}_1 \mat{Y}_{r_C} + \mat{B}_p\right]
\end{equation}
where the $4\times8$ projection matrix $\mat{P}$ is defined as,
\begin{equation}
P_{ij} = \delta_{i1}\delta_{j3} + \delta_{i2}\delta_{j4} + \delta_{i3}\delta_{j6},
\end{equation}
\noindent which extracts the third, fourth, and sixth rows of the four linearly independent solutions at the surface. Summing Equations \ref{eq:num_extract_porous} and \ref{eq:num_extract_surf} give the solutions that must be then constrained by the boundary conditions. This can be used to find the four unknown integration constants, $\vec{c}$,
\begin{equation}\label{eq:ode_find_c}
\Bigl[\left(\mat{P}\mat{D}_3 + \mat{P}_8\right) \mat{D}_2 \left(\mat{D}_1 \mat{I}_c + \mat{B_p}\right)\Bigr] \vec{c} = 	\vec{b} 
\end{equation}
\noindent which is a standard problem that can be solved through any appropriate inverse method.

The solution to the original coupled ODE system, Eq. \ref{eq:ode_diff_form}, is then given by multiplying the linearly independent solutions by the vector integration constant found by solving Equation \ref{eq:ode_find_c},
\begin{equation}
\vec{y}(r_{i,j}) = \mat{Y}(r_{i,j}) \vec{c}. 
\end{equation}

For more generalised interior structures, boundary conditions, and continuity conditions, the reader is referred to the excellent discussion in \Kporo, \S3.1.

Like \RNporo and \Kporo, our numerical procedure to solve the problem becomes unstable at small permeabilities/large melt viscosities. This instability is related to the decrease in thickness of the compaction boundary layer as $k/\eta_l$ decreases, which can be seen in Figure \ref{fig:heating_mobility}b. Stability can be somewhat increased with additional radial resolution, higher numerical precision, or by adopting the layer-splitting approaching outlined in \cite{kamata2023PoroviscoelasticGravitational}. For our purposes, we are interested in scenarios that maximise tidal dissipation. These require high mobilities and are thus generally more stable, so we do not try to develop an alternative workaround. Future work addressing this stability issue would be helpful.   

Our code that implements the above solution to the problem can be found at the GitHub page for \hyperlink{https://github.com/hamishHay/Love.jl}{Love.jl}.

\section{Tidal Forcing  and Potential Love Number}\label{app:forcing}

Io's diurnal tide is predominantly driven by Jupiter's tidal-forcing potential due to orbital eccentricity. In the reference frame rotating with Io with angular velocity $\omega = \Omega$, the degree-2 potential may be written as,
\begin{equation}
	\Phi^T_2 (t, r, \theta, \phi) = \s{\Phi}^T_{20}(r) P_{20}(\theta, \varphi) e^{i\Omega t} + \left[\s{\Phi}^{T,W}_{22}(r)  e^{i(2\varphi + \Omega t)} + \s{\Phi}^{T,E}_{22}(r) e^{i(2\varphi - \Omega t)} \right] P_{22}(\theta, \varphi) + c.c. 
\end{equation}
where the tidal forcing coefficients are $\f{\Phi}_{20}^T = -(3/2)\Omega^2 r^2 e$, $\f{\Phi}_{22}^{T,E} =(7/8)\Omega^2 r^2 e$, and $\f{\Phi}_{22}^{T,W} = -(1/8)\Omega^2 r^2 e$, and $c.c$ is the complex conjugate. 

The response of any body to a tidal-forcing is described by its potential Tidal Love number \cite{love1911ProblemsGeodynamics}. For a forcing at degree-$n$, this is given by the ratio of the response gravitational potential to the forcing potential,
\begin{equation}\label{eq:k2}
	k_n = \frac{(\s{\Phi}_1)_{nm}}{\s{\Phi}^T_{nm}}
\end{equation}
where $(\s{\Phi}_1)_{nm}$ is the complex degree-$n$ and order-$m$ response (i.e., perturbation) potential coefficient that arises from deformation, which is found by solving Eq. \ref{eq:perturb_pot_1}.

\section{Macro versus micro energy dissipation} \label{app:checks}

\begin{figure}[!h]
	\centering
	\includegraphics[width=0.8\linewidth]{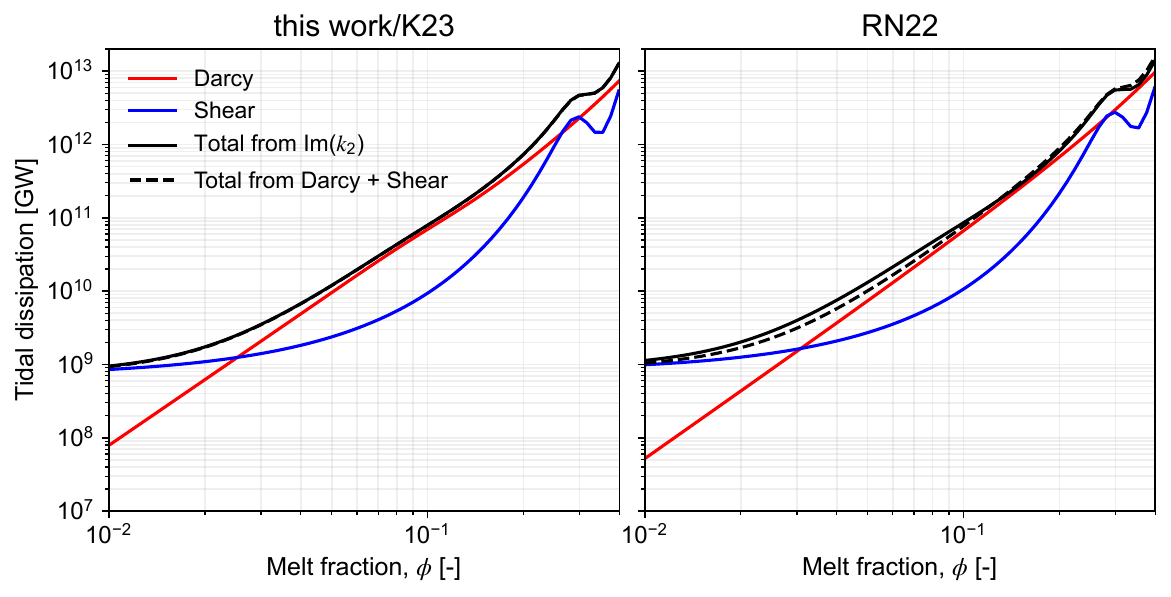}
	\caption{Macro versus micro energy dissipation calculation with the approach used in this work/\Kporo (left), and \RNporo (right). The black dashed lines (macro, Eq. \ref{eq:diss_E_k2}) and black solid lines (micro, Eq. \ref{eq:diss_def}) are indistinguishable in the left panel, but slightly different on the right.}
	\label{fig:K23_vs_RN22}
\end{figure}

There are differences in the implementation of the two-phase problem between \RNporo and \Kporo. As described in the main text, these difference stem from treatment of advection of the hydrostatic pressure and density fields. A consequence of this difference is that, following the method of \RNporo, the micro \eqref{eq:diss_def} and macro \eqref{eq:diss_E_k2} approaches of calculating energy dissipation do not give identical results. This difference is demonstrated in Figure \ref{fig:K23_vs_RN22}. We see that summing the individual heating contributions, (shear \eqref{eq:diss_sol_shear} and Darcy heating \eqref{eq:diss_liq_mobility}), gives identical results to using the $k_2$ Love number \eqref{eq:diss_E_k2} for the method used here and in \Kporo (left), but slightly different results for \RNporo (right). 

\printbibliography

\end{document}